\documentclass[sigconf]{acmart}
\AtBeginDocument{
  \providecommand\BibTeX{{
    \normalfont B\kern-0.5em{\scshape i\kern-0.25em b}\kern-0.8em\TeX}}}

\setcopyright{acmcopyright}
\copyrightyear{2018}
\acmYear{2018}
\acmDOI{10.1145/1122445.1122456}

\acmConference[Woodstock '18]{Woodstock '18: ACM Symposium on Neural
  Gaze Detection}{June 03--05, 2018}{Woodstock, NY}
\acmBooktitle{Woodstock '18: ACM Symposium on Neural Gaze Detection,
  June 03--05, 2018, Woodstock, NY}
\acmPrice{15.00}
\acmISBN{978-1-4503-XXXX-X/18/06}
\usepackage{subfigure,epsfig}
\usepackage{algorithm}
\usepackage{algorithmic}
\usepackage{multirow}
\usepackage{float}
\usepackage{url}
\usepackage{amsthm,amsmath}

\usepackage{mathrsfs}

\usepackage{enumitem}

\usepackage{array}
\usepackage{booktabs}
\renewcommand\arraystretch{0.9}
\setlist[itemize]{leftmargin=*}
\setlist[enumerate]{leftmargin=*}

\newcommand{\Mat}[1]{\mathbf{#1}}

\newcommand{\Set}[1]{\mathcal{#1}}

\newcommand{\ie}{\emph{i.e., }}
\newcommand{\eg}{\emph{e.g., }}

\def\BibTeX{{\rm B\kern-.05em{\sc i\kern-.025em b}\kern-.08emT\kern-.1667em\lower.7ex\hbox{E}\kern-.125emX}}

\begin{document}

\title{Contrastive Learning for Cold-Start Recommendation}

\author{Yinwei Wei}
\email{weiyinwei@hotmail.com}
\affiliation{\institution{National University of Singapore}
\country{}}

\author{Xiang Wang}
\email{xiangwang@u.nus.edu}
\affiliation{\institution{National University of Singapore}\country{}}

\author{Qi Li}
\email{iliqi@mail.sdu.edu.cn}
\affiliation{\institution{Shandong University}\country{}}

\author{Liqiang Nie}
\email{nieliqiang@gmail.com}
\affiliation{\institution{Shandong University}\country{}}

\author{Yan Li}
\email{liyan@kuaishou.com}
\affiliation{\institution{Beijing Kuaishou Technology Co., Ltd.}\country{}}

\author{Xuanping Li}
\email{lixuanping@kuaishou.com}
\affiliation{\institution{Beijing Kuaishou Technology Co., Ltd.}\country{}}

\author{Tat-Seng Chua}
\email{chuats@comp.nus.edu.sg}
\affiliation{\institution{National University of Singapore}\country{}}

\begin{abstract}
Recommending purely cold-start items is a long-standing and fundamental challenge in the recommender systems.
Without any historical interaction on cold-start items, the collaborative filtering (CF) scheme fails to leverage collaborative signals to infer user preference on these items.  
To solve this problem, extensive studies have been conducted to incorporate side information of items (\eg content features) into the CF scheme. 
Specifically, they employ modern neural network techniques (e.g., dropout, consistency constraint) to discover and exploit the coalition effect of content features and collaborative representations. 
However, we argue that these works less explore the mutual dependencies between content features and collaborative representations and lack sufficient theoretical supports, thus resulting in unsatisfactory performance on cold-start recommendation.

In this work, we reformulate the cold-start item representation learning from an information-theoretic standpoint. 
It aims to maximize the mutual dependencies between item content and collaborative signals. 
Specifically, the representation learning is theoretically lower-bounded by the integration of two terms: mutual information between collaborative embeddings of users and items, and mutual information between collaborative embeddings and feature representations of items. 
To model such a learning process, we devise a new objective function founded upon contrastive learning and develop a simple yet effective Contrastive Learning-based Cold-start Recommendation framework (CLCRec). 
In particular, CLCRec consists of three components: contrastive pair organization, contrastive embedding, and contrastive optimization modules. 
It allows us to preserve collaborative signals in the content representations for both warm and cold-start items. Through extensive experiments on four publicly accessible datasets, we observe that CLCRec achieves significant improvements over state-of-the-art approaches in both warm- and cold-start scenarios.
\end{abstract}


\maketitle

\section{Introduction}
Collaborative Filtering (CF) has exhibited remarkable success in personalized recommendation.
Learning quality representations of users and items from historical interactions (\eg clicks, rates, views) is central to modern CF models~\cite{BPR,NCF,LightGCN,MMGCN}.
We denote the representations encoding collaborative signals as collaborative embeddings, such as ID embeddings in MF \cite{BPR}, interaction history embeddings in SVD++ \cite{SVD++}, and interaction graph embeddings in LightGCN \cite{LightGCN}.
Such collaborative embeddings reflect behavioral similarity of users, thus inferring user preferences on items effectively.
Clearly, historical user-item interactions are of crucial importance to establish high-quality collaborative embeddings.
However, there are a large amount of newly coming items, which never appear in the history, posing the serious challenge of cold-start recommendation.
This is because unavailable interaction fails the learning of collaborative embedding for these cold-start items.

To address this problem, one common solution is incorporating item content (\eg images, videos, attributes) into the CF scheme.
The content features are supposed to cooperate with collaborative representations, form the coalition, and offer better predictions.
Such coalition effects memorized in CF models allow us to transfer collaborative knowledge to cold-start items.
In this research line, extensive studies have been conducted, which can be roughly categorized into two types, robustness-based \cite{DropoutNET,MTPR,CCCC,WarmUp} and constraint-based \cite{LLAE,LCE,CB2CF,Decoupled,Heater}.
Despite success, each type of methods suffers from some inherent limitations:
\begin{itemize}[leftmargin=*]
    \item From the viewpoint of robust learning, robustness-based methods frame the cold-start items as the corrupted forms of warm items whose interaction histories are missing.
    Technically, they augment the training data by randomly corrupting warm items' collaborative embeddings~\cite{DropoutNET, MTPR, CCCC} or manually crafting new items~\cite{Meta,MeLU,MHI,WarmUp}, so as to enhance the generalization ability to unseen items.
    However, no function is designed to explicitly consider the relationships between item content and collaborative embeddings, thereby hardly refining useful collaborative signals for cold-start items.
    \item Constraint-based methods~\cite{CB2CF,Decoupled,Heater} explicitly model the relationships between item content and collaborative embeddings by applying a constraint loss.
    Specifically, for an item, its content is transferred as feature representation;
    thereafter, a constraint loss (\eg $L_{2}$ regularizer) enforces dimension-wise similarity between feature representation and collaborative embedding.
    Nonetheless, there are two limitations:
    (1) such dimension-wise regularization merely remains the consistency between these two representations, which possibly discards the heterogeneous and unique information;
    (2) it lacks sufficient theoretical support to ensure what and how much information on collaborative embeddings is preserved in feature representations.
\end{itemize}
In a nutshell, while being powered by the advance of neural network techniques (\eg dropout~\cite{dropout} and constraint~\cite{Heater}), these existing methods are insufficient to boost cold-start item recommendation.

In this work, we aim to explore the in-depth and theoretical aspect of cold-start representation learning.
Specifically, we reformulate the learning of feature representation as optimizing its maximum posterior estimator under the observed interactions and learned collaborative embeddings.
This optimization allows us to distill meaningful information from collaborative embeddings, and encode them into the feature representations for both warm and cold-start items.
More formally, the lower bound of the posterior estimator is the integration of two terms:
(1) \textbf{mutual information between user and item collaborative embeddings} (short for U-I) and (2) \textbf{mutual information between feature representation and collaborative embedding} (short for R-E).
Towards the maximum of mutual information, we propose a novel objective function founded upon contrastive learning~\cite{CPC,SimCLR,Infoxlm,CMC} from an information-theoretic standpoint.

To achieve our proposed objective function, we further develop a general framework, named \emph{Contrastive Learning-based Cold-start Recommendation} (CLCRec).
Technically, it consists of three key components.
(1) We first conduct \textbf{contrastive pair organization}. For the U-I loss, we treat observed historical interactions as positive user-item pairs, while unobserved interactions as negative user-item pairs.
Meanwhile, for the R-E loss, we adopt the self-discrimination task of items, where one item paired with itself and other items refer to the positive and negative item-item pairs, respectively.
(2) Having established the U-I and R-E contrastive pairs, we build \textbf{contrastive embedding networks} (CEN) upon them.
The U-I CEN is to generate the collaborative embeddings for the user-item pairs, which can be built upon any CF encoder like MF and LightGCN.
The R-E CEN contains a trainable encoder to learn feature representations of items based on their content features.
(3) Afterwards, we perform \textbf{contrastive optimization} according to our designed objective function.
Using the pre-defined density ratio functions, we score the constructed U-I and R-E contrastive pairs and then perform contrastive training --- that is, identifying the positive pair from multiple negative pairs.
As such, two CENs are trained to guide the learning of feature representations and collaborative embeddings, and encourage the information transfer between them.
In the testing phase, for a cold-start item, we are able to generate high-quality feature representations that memorize informative signals from observed interactions, and predict how likely a user would adopt it.
Through extensive experiments on four real-world datasets, we demonstrate that CLCRec achieves significant improvements over the state-of-the-art approaches in both warm- and cold-start scenarios. 
We summarize the contributions: 
\begin{itemize}
    \item We reformulate the cold-start representation learning as maximizing the sum of U-I and R-E mutual information.
    \item Towards mutual information maximum, we design a contrastive learning-based objective function and accordingly propose a general framework, CLCRec. 
    \item We conduct extensive experiments on four public datasets to demonstrate the superiority of CLCRec. We have released the codes and datasets: \url{https://github.com/weiyinwei/CLCRec} to facilitate reproducibility.
\end{itemize}

\section{Methodology}
Generally, the cold-start problem explores how to predict the affinity of new items, which has no or few interaction records with users in the training set.
In this work, we focus on recommending the complete cold-start items without any interactions, which is more challenging and promising. 
Before describing the detail of the proposed method, we reformulate the cold-start problem and accordingly derive the objective function.

\subsection{Problem Formalization}
\noindent\textbf{Learning of Collaborative Embeddings.}
Let $\mathcal{O}=\{o_{u,i}\}$ be the set of user-item interactions, which involves the set of $N$ users, $\Set{U}=\{u\}$ and the set of $M$ items, $\Set{I}=\{i\}$.
With the assumption that users with similar behavior prefer similar items, modern collaborative filtering (CF) methods \cite{LF1,JH,LF2} learn collaborative embeddings of users and items from the historical interactions.
Without loss of generality, we describe an user $u$ (a item $i$) with a collaborative embedding $\mathbf{z}_{u}\in\mathbb{R}^d$ ($\mathbf{z}_{i}\in\mathbb{R}^d$), where $d$ is the embedding size.
Optimizing collaborative embeddings is to maximize the posterior estimator given the observed interactions:
\begin{equation}
    \Mat{Z}^{*} = \arg\max_{\mathbf{Z}}p(\Mat{Z}|\Set{O}),
\end{equation}
where $\Mat{Z} = [\Mat{z}_{i_1},\cdots,\Mat{z}_{i_N},\ \Mat{z}_{u_1},\cdots,\Mat{z}_{u_M}]$ collects the collaborative embeddings of all users and items; $p(\Mat{Z}|\Set{O})$ is to encode as much information on $\Set{O}$ into $\Mat{Z}$ as possible.
However, for cold-start items $\Set{I}^{c}=\{i^c\}$ without any historical interactions, CF is infeasible to optimize their collaborative embeddings, thus fails to predict user preference accurately.

\vspace{5pt}
\noindent\textbf{Learning of Feature Representations.}
To alleviate this problem, extensive studies have been conducted to exploit the rich side information of items, especially content features (\eg frames of video, soundtracks of music, and descriptions of product).
In general, a feature encoder $f_{\theta}$ is built upon the content feature vector $\Mat{x}_i$ to generate the feature representation $\Mat{f}_i$ of item $i$.
Here we establish the collections of feature representations and content features as $\Mat{F}=[\Mat{f}_{i_{1}},\cdots,\Mat{f}_{i_{N}}]$ and $\Mat{X}=[\Mat{x}_{i_{1}},\cdots,\Mat{x}_{i_{N}}]$, respectively. 
Note that, $\Mat{F}$ that are learnable, thus differs from $\Mat{X}$ that are learned by pre-trained feature extractors~\cite{LM1, wei2019neural}. 
Previous works either treat the feature representation and collaborative embeddings individually \cite{DropoutNET,MTPR,CCCC,WarmUp} or apply a regularizer term to encourage their dimension-wise similarity \cite{LLAE,LCE,CB2CF,Decoupled,Heater}.

Unlike these methods, we optimize the feature encoder by encouraging the feature representations to (1) encode collaborative signals from collaborative embeddings and (2) preserve affinities with users who have interacted with before.
To this end, we reformulate the item representation learning as maximizing the posterior estimator, given the embeddings and interactions:
\begin{equation}\label{equ:overall-optimization}
   \theta^{*} = \arg\max_{\theta}p(\theta|\Mat{Z},\Set{O},\Mat{X}).
\end{equation}
Having obtained the optimized feature encoder $\theta^{*}$, we generate feature representations for cold-start items, which is enhanced with the collaborative knowledge from warm items. 
Finally, we could infer the interaction probability between a cold-start item and a user by measuring the similarities of their representations.

\subsection{Objective Function}
As the pre-existing content features $\Mat{X}$ are usually independent to the collaborative embeddings $\Mat{Z}$, we can adopt Bayesian rules and rewrite Equation \eqref{equ:overall-optimization} as follows:
\begin{gather}
    p(\theta|\Mat{Z},\Set{O},\Mat{X})\propto p(\Mat{Z},\Set{O}|\Mat{F})\cdot p(\theta).
\end{gather}
Following the prior study~\cite{BPR}, we assume that all items (and users) are independent with each other.
As such, we can further obtain the following probability to associate the feature representation with the collaborative embeddings and interactions, respectively:
\begin{align}
    &p(\Mat{Z},\Set{O}|\Mat{F})\cdot p(\theta)\nonumber
    \propto  \sum_{i\in\Set{I}}\ln{p(\Mat{z}_i,\Mat{z}_{u_1}, \Mat{z}_{u_2},\dots,\Mat{z}_{u_{N}} | \Mat{f}_i)}+\ln{p(\theta)}\\
    =& \sum_{i\in\Set{I}}\ln{p(\Mat{z}_i, \Mat{f}_i)}+\sum_{(u,i)\in\Set{O}}\ln{p(\Mat{z}_u|\Mat{z}_i, \Mat{f}_i)}+C\sum_{u\in\Set{U}}\ln{p(\Mat{z}_u)}+\ln{p(\theta)}.\nonumber
\end{align}
Here $C$ is a constant term derived from the sum of conditional probabilities: $\sum_{(u,i)\not\in\Set{O}}\ln{p(\Mat{z}_u|\Mat{z}_i, \Mat{f}_i)}$.
It captures the intuition that an item is independent of the users who did not interact with before, whereas the presence of an interaction suggests their independence.

In this equation, the first term focuses on the correlation between $\Mat{z}_i$ and $\Mat{f}_i$, such that $\Mat{z}_i$ is encouraged to embed the information of $\mathbf{f}_i$.
Thus, we could relieve the second term $\sum_{(u,i)\in\Set{O}}\ln{p(\Mat{z}_u|\Mat{z}_i, \Mat{f}_i)}$ from the complex relationship among three kinds of representations.
Specifically, it is proportional to the probabilities of user embedding, after observing the item embedding:
\begin{align}
      &\sum_{i\in\Set{I}}\ln{p(\Mat{z}_i, \Mat{f}_i)}+\sum_{(u,i)\in\mathcal{O}}\ln{p(\mathbf{z}_u|\mathbf{z}_i, \mathbf{f}_i)}+C\sum\limits_{u\in\mathcal{U}}\ln{p(\mathbf{z}_u)}+\ln{p(\theta)}\nonumber\\
    \propto&\sum\limits_{i\in\mathcal{I}}\ln{p(\mathbf{z}_i, \mathbf{f}_i)}+\sum\limits_{(u,i)\in\mathcal{O}}\ln{p(\mathbf{z}_u|\mathbf{z}_i)}+C\sum\limits_{u\in\mathcal{U}}\ln{p(\mathbf{z}_u)}+\ln{p(\theta)}.\nonumber
\end{align}

Afterwards, we prove that the lower bound of this posterior probability is the sum of two components: (1) U-I mutual information between collaborative embeddings of users and items, and (2) R-E mutual information between collaborative embeddings and feature representations of items:
\begin{align}
    &\sum\limits_{i\in\mathcal{I}}\ln{p(\mathbf{z}_i, \mathbf{f}_i)}+\sum\limits_{(u,i)\in\mathcal{O}}\ln{p(\mathbf{z}_u|\mathbf{z}_i)}+C\sum\limits_{u\in\mathcal{U}}\ln{p(\mathbf{z}_u)}+\ln{p(\theta)}\nonumber\\
    \geq &\underbrace{\sum_{i\in\mathcal{I}} p(\mathbf{f}_i, \mathbf{z}_i)\ln\frac{p(\mathbf{z}_i, \mathbf{f}_i)}{p(\mathbf{z}_i)p( \mathbf{f}_i)}}_{MI(\Mat{F}, \Mat{Z}_I)}+ \underbrace{\sum_{(i,j)\in\mathcal{O}}  p(\mathbf{z}_u, \mathbf{z}_i) \ln\frac{p(\mathbf{z}_u|\mathbf{z}_i)}{p(\mathbf{z}_u)}}_{MI(\Mat{Z}_I,\Mat{Z}_U)}\nonumber\\
    + &\underbrace{\sum\limits_{i\in\mathcal{I}}\ln{p(\mathbf{f}_i)}+\sum\limits_{i\in\mathcal{I}}\ln{p(\mathbf{z}_i)}+(C+1)\sum\limits_{u\in\mathcal{U}}\ln{p(\mathbf{z}_u)}+\ln{p(\theta)}}_{\ln{p(\Theta)}}\nonumber
\end{align}
where $MI(\cdot,\cdot)$ represents the mutual information between two representations. In addition, we use $\Theta$ to represent all the parameters whose prior probabilities presumably follow Gaussian Distributions. 
Therefore, we cast the objective from the maximum posterior estimator to mutual information maximization. 

\begin{figure}
\centering
\includegraphics[scale=0.45]{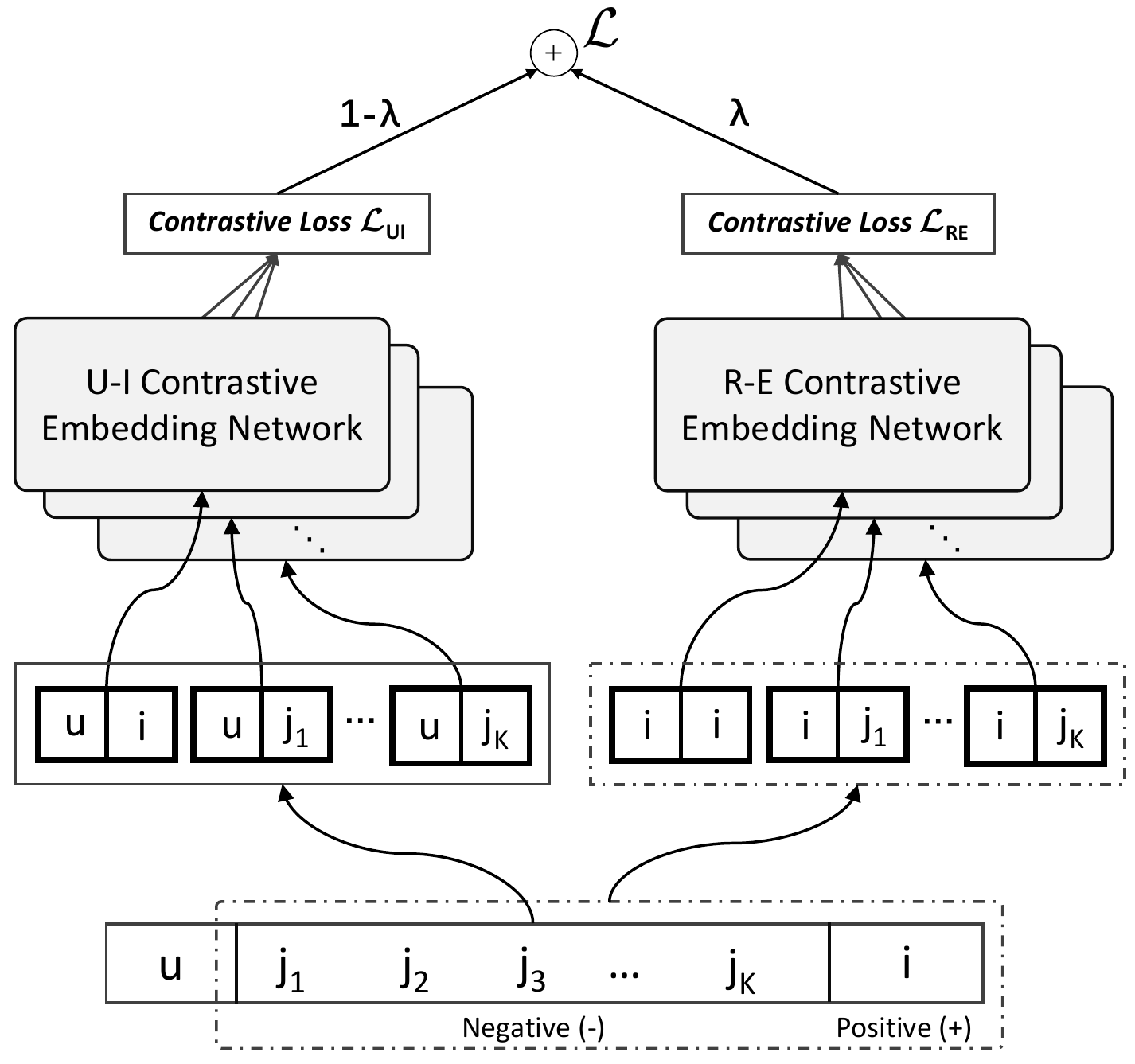}
\caption{Schematic illustration of our proposed framework. It illustrates the contrastive pair organization, contrastive embedding network, and contrastive optimization.}
\label{framework}
\vspace{-10pt}
\end{figure}

To maximize the mutual information, we employ the technique of contrastive learning, which models the mutual dependencies of variable pairs by recognizing the positive pair from the negative ones. 
Specifically, following the prior work~\cite{CPC}, we define two density ratio functions: 
\begin{equation}
    h(\mathbf{z}_i, \mathbf{f}_i)\propto\frac{p({\mathbf{f}_i|\mathbf{z}_i})}{p(\mathbf{f}_i)}, \ \ \  
    g(\mathbf{z}_u, \mathbf{z}_i)\propto\frac{p(\mathbf{z}_u|\mathbf{z}_i)}{p(\mathbf{z}_u)}.
    \label{eq6}
\end{equation}
These functions~(\textit{i.e.,} $h(\cdot,\cdot)$ and $g(\cdot,\cdot)$) could be used to measure the correlations between two variables. 
As a result, the contrastive learning-based objective function could be formulated as:
\begin{align}
    \mathcal{L}=&\lambda\mathcal{L}_{RE} + (1-\lambda)\mathcal{L}_{UI} + \eta||\Theta||_2^2\\
    =&-\lambda\underset{i\in\mathbf{I}}{\mathbb{E}}\Big[\ln\frac{h(\mathbf{z}_i, \mathbf{f}_i)}{h(\mathbf{z}_i, \mathbf{f}_i)+\sum\limits_{j\in \mathbf{I}/\{i\}}{h(\mathbf{z}_i, \mathbf{f}_j)}}\Big]\nonumber\\
    &- (1-\lambda)\underset{(u,i)\in\mathcal{O}}{\mathbb{E}}\Big[\ln\frac{g(\mathbf{z}_u, \mathbf{z}_i)}{g(\mathbf{z}_u, \mathbf{z}_i)+\sum\limits_{(u,j)\notin\mathcal{O}}{g(\mathbf{z}_u, \mathbf{z}_j)}}\Big] + \eta||\Theta||_2^2.\nonumber
\end{align}
We use $\mathcal{L}_{RE}$ and $\mathcal{L}_{UI}$ to denote the contrastive loss for maximizing the R-E and U-I mutual information, respectively. 
As $\mathcal{L}_{UI}$ has the same optimal objective with Personalized Ranking~(BPR) loss \cite{BPR}, we could optimize the recommendation model in an end-to-end manner without any other loss function, and set a hyper-parameter $\lambda$ to balance the collaborative and feature representations learning.

\subsection{CLCRec}
According to the objective function, we develop a general framework, Contrastive Learning-based Cold-start Recommendation (CLCRec), as shown in Figure~\ref{framework}. This framework is composed of three components: contrastive pair organization, contrastive embedding network, and contrastive optimization. 
In what follows, we elaborate the detail of each component.

\subsubsection{\textbf{Contrastive Pair Organization}}
The core of contrastive learning is to identify the positive pair constructed by semantic similar instances from some negative ones paired by the dissimilar samples. 
As such, in this work, it is critical to organize the users and items as U-I and R-E contrastive pairs. 

\vspace{5pt}
\noindent\textbf{U-I Contrastive Pair. } 
We view the user-item pair observed in the historical interactions as the positive sample, like $(u, i)$ shown in Figure~\ref{framework}. 
Meanwhile, we randomly sample $K$ items (\textit{e.g.,} $(j_{1}, j_2,\cdots, j_K)$), which have not been purchased by $u$, and pair the user to establish the negative pairs, as illustrated in Figure~\ref{framework}. 
Formally, the positive and negative U-I pairs can be defined as, 
\begin{equation}
    \{(u, i), (u, j_1),(u, j_2),\cdots,(u,j_K)\}.
\end{equation} 
Compared with the negative pairs, the positive one contains the similar collaborative signal. Hence, such a comparison facilitates the discovery of collaborative signal conveyed by the interactions.

\vspace{5pt}
\noindent\textbf{R-E Contrastive Pair. }
Different from U-I pairs, R-E pairs adopt the self-discrimination task to maximize the mutual information of two different representations of items. 
To construct the pairs, taking $i$ in Figure~\ref{framework} as an example, we set it as the anchor item and join the anchor with itself as the positive pair, which reveals the semantic similarity between two representations of the same item. 
By contrast, the negative pairs organized by the anchor with other items are semantically dissimilar. 
As shown in Figure~\ref{framework}, we pair the items together to obtain the R-E pairs:
\begin{equation}
    \{(i,i), (i,j_1), (i,j_2), \cdots, (i,j_K)\},
\end{equation}
where $(i, i)$ is the positive pair and the others are negative ones. 

\begin{figure}
\centering
\includegraphics[scale=0.5]{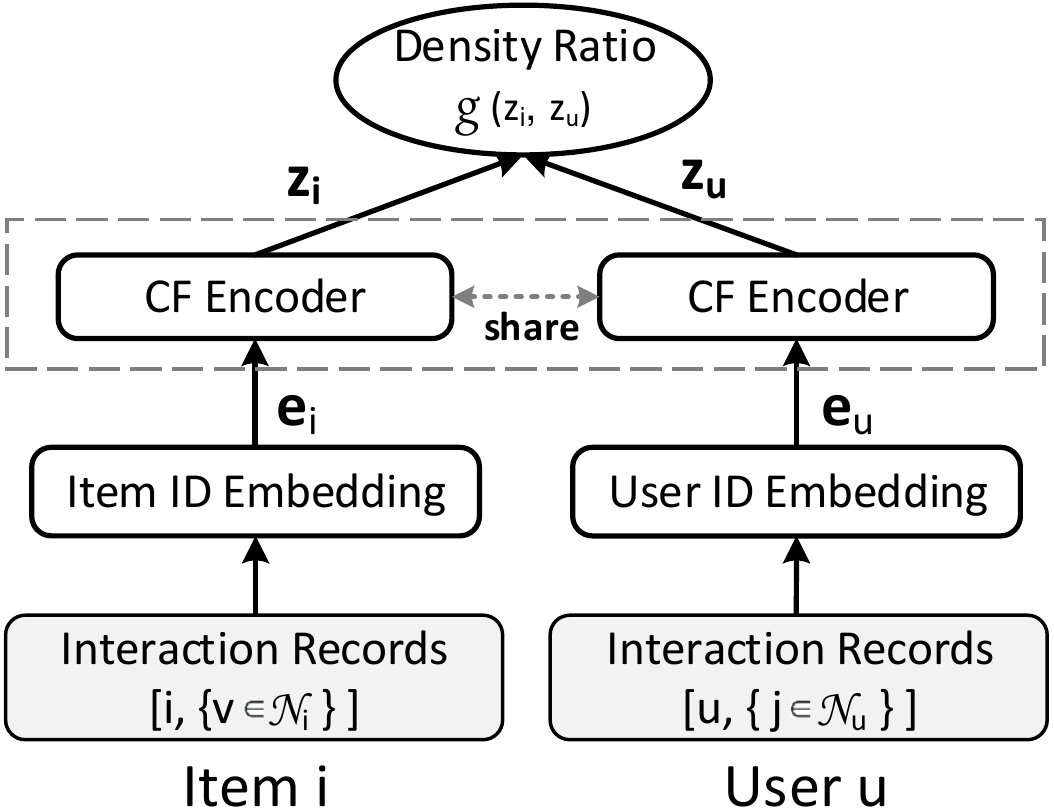}
\caption{Illustration of U-I CEN.}
\label{ui_ce}
\vspace{-5pt}
\end{figure}
\subsubsection{\textbf{Contrastive Embedding Network}}
Upon contrastive pairs, we devise the U-I and R-E contrastive embedding networks (CENs) to represent the user and item and calculate the correlation of each pair using the pre-defined density ratio functions. 

\vspace{5pt}
\noindent \textbf{U-I Contrastive Embedding Network. } 
To model collaborative signal from user-item interactions, we first fetch their id embeddings~(\textit{e.g.,} $\textbf{e}_u$ and $\textbf{e}_i$) from an look-up table, which is defined by a parameter matrix:
\begin{equation}
    \mathbf{E}=[\mathbf{e}_{i_1}, \dots, \mathbf{e}_{i_N}, \mathbf{e}_{u_1}, \dots, \mathbf{e}_{u_M}].
\end{equation}
 
Then, a shared CF encoder is utilized to learn the collaborative embeddings for the users and items, as shown in Figure~\ref{ui_ce}. 
It can be implemented by various models, such as MF-based~\cite{MF,PMF,BPR}, neural network-based~\cite{NCF}, and graph neural network-based~\cite{NGCF,LightGCN,wei2021hierarchical} models. 
In this work, we provide two simple yet efficient implementations, $CEN_{MF}$ and $CEN_{GCN}$, which are based on MF~\cite{PMF} and LightGCN~\cite{LightGCN}, respectively. 
For conciseness, we neglect their details and formulate them as,
\begin{equation}
\begin{cases}
    \mathbf{z}_i = \mathcal{E}(\mathbf{e}_i, \{\mathbf{e}_{v}|v\in\Set{N}_{i}\})\\
    \mathbf{z}_u = \mathcal{E}(\mathbf{e}_u, \{\mathbf{e}_{j}|j\in\Set{N}_{u}\}),
\end{cases}
\end{equation}
where, $\mathcal{E}$ is the CF encoder taking id embeddings as input; $\Set{N}_{i}$ is the set of users who have interacted with item $i$, while $\Set{N}_{u}$ represents the set of items which have been purchased by user $u$.
Having obtained the collaborative embeddings of user-item pair (\ie $\mathbf{z}_i$ and $\mathbf{z}_u$), we define the following function to measure their correlation:
\begin{equation}
    g(\mathbf{z}_i, \mathbf{z}_u) = \exp(\mathbf{z}_i^\intercal\mathbf{z}_u/\tau),
\end{equation}
where $\tau$ is a temperature parameter~\cite{SimCLR}. 
Here we employ the inner product to approach the correlation, leaving the exploration of other non-negative functions \cite{CPC} in future work.

\begin{figure}
\centering
\includegraphics[scale=0.5]{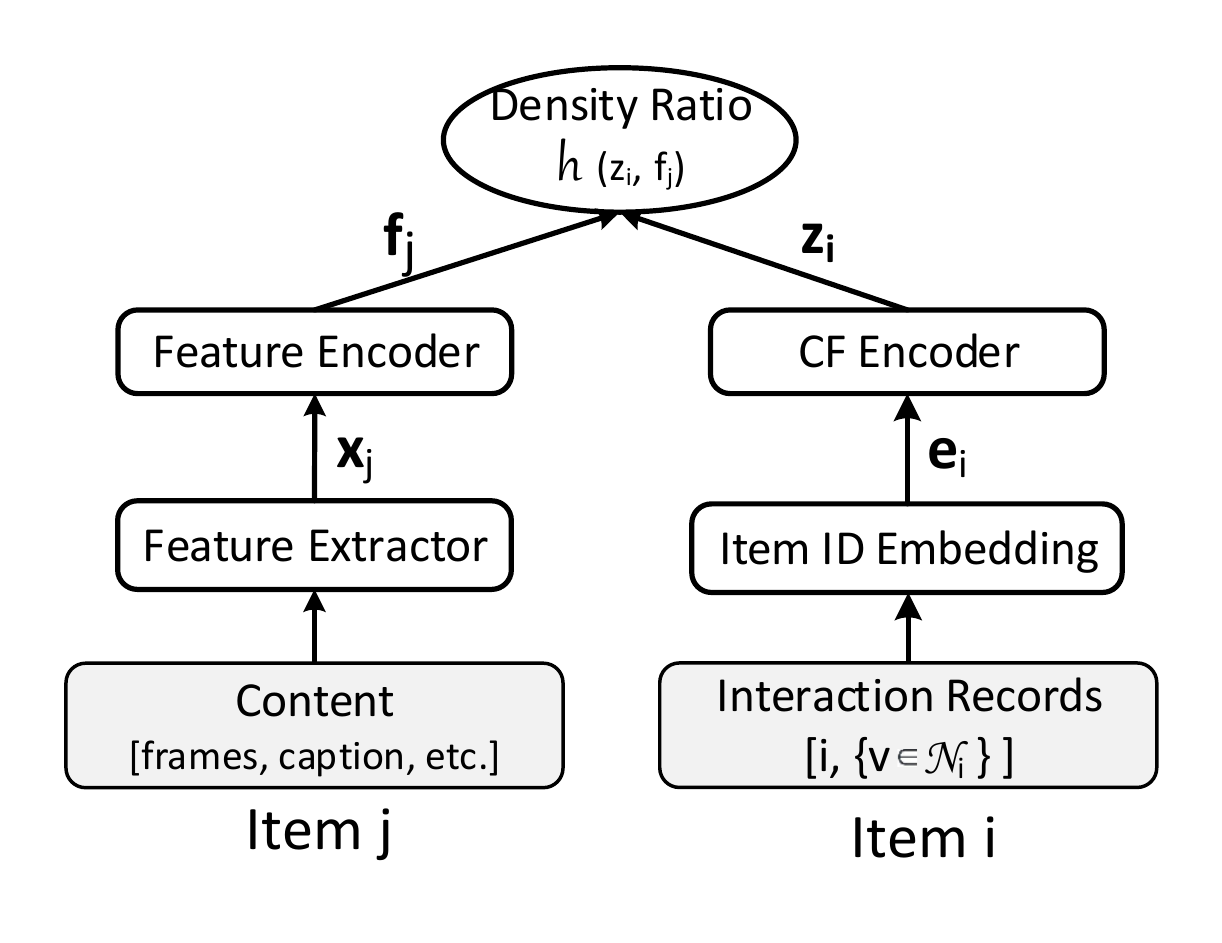}
\caption{Illustration of R-E CEN.}
\label{cf_ce}
\vspace{-5pt}
\end{figure}

\vspace{5pt}
\noindent\textbf{R-E Contrastive Embedding Network. }
This network is used to score the correlation between two views of items --- collaborative signals and content information. 
As shown in Figure~\ref{cf_ce}, it can be regarded as two parallel pipelines to model the item's feature representation and collaborative embedding, respectively. 
The right one for the collaborative embedding is the same with the corresponding operations in the U-I network. 

Towards representing the content information, we utilize the feature extractor, which could be a pre-trained deep learning model, followed by the feature encoder~\cite{JI1}. 
More specifically, for item $j$ in the R-E contrastive pair, we extract its feature vector, denoted as $\mathbf{x}_j$, from the content information. 
We then design the feature encoder as a multiple layer perceptron~(MLP) with one hidden layer, formally 
\begin{equation}
    \mathbf{f}_j = \mathbf{W}^{(2)}\phi(\mathbf{W}^{(1)}\mathbf{x}_j+\mathbf{b}^{(1)})+\mathbf{b}^{(2)},
\end{equation}
where $\mathbf{W}^{(\cdot)}$ and $\mathbf{b}^{(\cdot)}$ denote the trainable matrix and bias vector of the encoder, respectively. 
Moreover, $\mathbf{f}_j$ is the desired feature representation of item $j$, which aims to distill the content feature and maximumly preserve the information associating to the collaborative signal. 
In the inference phase, powered by such feature representations, we could measure the affinity between the user and cold-start item.

After obtaining the outputs of two pipelines, \textit{i.e.,} the item's collaborative embedding and feature representation, we score their density ratios with the following function:
\begin{equation}
    h(\mathbf{z}_i, \mathbf{f}_j) = \exp(\frac{\mathbf{z}_i^\intercal \mathbf{f}_j}{||\mathbf{z}_i||\cdot||\mathbf{f}_j||}\cdot\frac{1}{\tau}).
\end{equation}
We empirically normalize the vectors to measure the ratio in an unit space instead of projecting them into a common space, which helps to maintain the characters of content information. 
\subsubsection{\textbf{Contrasitve Optimization}}
In order to maximize the mutual information, we perform the contrastive training strategy to optimize the parameters of the model. 
Incorporating the defined density ratio functions, we implement the objective function as,
\begin{align}
    \mathcal{L}&=\lambda\mathcal{L}_{RE} + (1-\lambda)\mathcal{L}_{UI} + \eta||\Theta||_2^2\\
    &=-\lambda\underset{i\in\mathbf{I}}{\mathbb{E}}\Big[\ln\frac{\exp(\frac{\mathbf{z}_i^\intercal\,\mathbf{f}_i}{||\mathbf{z}_i||\cdot||\mathbf{f}_i||}\cdot\frac{1}{\tau})}{\exp(\frac{\mathbf{z}_i^\intercal\,\mathbf{f}_i}{||\mathbf{z}_i||\cdot||\mathbf{f}_i||}\cdot\frac{1}{\tau})+\sum_{k=1}^{K}{\exp(\frac{\mathbf{z}_i^\intercal\,\mathbf{f}_{j_k}}{||\mathbf{z}_i||\cdot||\mathbf{f}_{j_k}||}\cdot\frac{1}{\tau})}}\Big]\nonumber\\
    &-(1-\lambda)\underset{(u,i)\in\mathcal{O}}{\mathbb{E}}\Big[\ln\frac{\exp(\frac{\mathbf{z}_i^\intercal\mathbf{z}_u}{\tau})}{\exp(\frac{\mathbf{z}_i^\intercal\mathbf{z}_u}{\tau})+\sum_{k=1}^K{\exp(\frac{\mathbf{z}_{j_k}^\intercal\mathbf{z}_u}{\tau})}}\Big] +\eta||\Theta||_2^2.\nonumber
\end{align}

\begin{figure}
\centering
\includegraphics[scale=0.5]{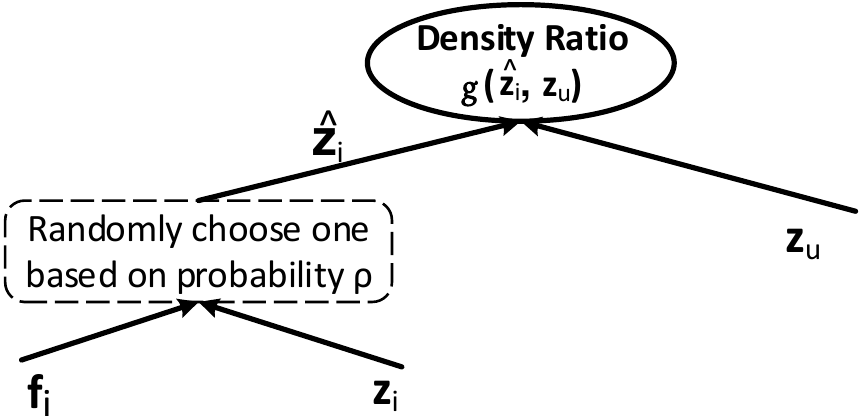}
\vspace{-5pt}
\caption{Illustration of the hybrid contrastive training.}
\label{random}
\vspace{-10pt}
\end{figure}
\noindent\textbf{Relation with BPR Loss. } Within the function, we find that BPR loss can be regarded as a special case of $\mathcal{L}_{UI}$, when setting $\tau$ as 1 and sampling one negative pair for training. 
However, we suggest that $\mathcal{L}_{UI}$ is superior to BPR loss. 
On one hand, it is intuitively closer to the real-world scenario that we tend to face multiple options instead of the either-or situation. 
On the other hand, we indicate the function intrinsically facilitates the model optimization. 
Specially, we calculate the gradient of $\mathcal{L}_{UI}$ with respect to $\mathbf{z}_u$ and $\mathbf{z}_i$:
\begin{equation}
      \frac{\partial \mathcal{L}_{UI}}{\partial \mathbf{z}_u} = \Big(\frac{\sum\limits_{(u,j)\notin\mathcal{O}}\exp({\mathbf{z}_j^\intercal\mathbf{z}_u/\tau})}{\exp(\mathbf{z}_i^\intercal\mathbf{z}_u/\tau)+{\sum\limits_{(u,j)\notin\mathcal{O}}\exp({\mathbf{z}_j^\intercal\mathbf{z}_u/\tau})}{}}\Big)\Big(\frac{\partial \mathbf{z}_j^\intercal}{\partial \mathbf{z}_u}-\frac{\partial \mathbf{z}_i^\intercal}{\partial \mathbf{z}_u}\Big)\frac{\mathbf{z}_u}{\tau},\nonumber
\end{equation}
\begin{equation}
      \frac{\partial \mathcal{L}_{UI}}{\partial \mathbf{z}_i} = \Big(\frac{\sum\limits_{(u,j)\notin\mathcal{O}}\exp({\mathbf{z}_j^\intercal\mathbf{z}_u/\tau})}{\exp(\mathbf{z}_i^\intercal\mathbf{z}_u/\tau)+{\sum\limits_{(u,j)\notin\mathcal{O}}\exp({\mathbf{z}_j^\intercal\mathbf{z}_u/\tau})}{}}\Big)\Big(-\frac{\mathbf{z}_i}{\tau}\Big)\frac{\partial \mathbf{z}_u^\intercal}{\partial \mathbf{z}_i}.\nonumber
      \label{eq_mg}
\end{equation}

During the training process, the negative items are increasingly pushed away from the users who have no interactions. 
Thus, the scores of the negative pairs are close to zero and the contribution of the above gradients comes mainly from hard negative items. 
For BPR loss, it is negatively influenced by the randomly negative sampling, which gains the hard negative in a low probability.  
Nevertheless, $\mathcal{L}_{UI}$ is benefited by its multiple negative samples, which obviously increases the probability of hard negative detection, thus offers meaningful and larger gradients to the optimization \cite{DBLP:conf/wsdm/RendleF14}.  

\vspace{5pt}
\noindent\textbf{Hybrid Contrastive Training. }
As mentioned in Section 2.2, maximizing the R-E mutual information improves the correlation between $\mathbf{z}_i$ and $\mathbf{f}_i$, and then approximates $\sum_{(u,i)\in\mathcal{O}}\ln{p(\mathbf{z}_u|\mathbf{z}_i, \mathbf{f}_i)}$ to $\sum_{(u,i)\in\mathcal{O}}\ln{p(\mathbf{z}_u|\mathbf{z}_i)}$. 
However, there is still a gap between them at the beginning of the training phase. 
To alleviate this drawback, we empirically utilize a hybrid training strategy in the U-I CEN, as shown in Figure~\ref{random}. 
In particular, we randomly choose the learned feature representation or item collaborative embedding with a pre-defined probability $\rho$ and feed it into the density ratio function. It is able to maximize the correlation between learned features and user preference as well as model the collaborative embeddings in the training phase. 
Taking the advantage of this design, we could predict the affinities between users and cold-start items by measuring the similarities of their representations in the inference phase.
\vspace{-3pt}
\section{Experiments}
In this section, we first introduce the datasets for evaluation and describe the experiment setting. 
Then, to evaluate the effectiveness of our proposed method, we compare our proposed method with the baselines in the warm-start, cold-start, and all item scenarios. 
Moreover, we analyze the designs of our proposed framework through an ablation study. 
\vspace{-3pt}
\subsection{Experimental Settings}
\label{exp_set}
\subsubsection{\textbf{Dataset}}
To evaluate the effectiveness of CLCRec, extensive experiments are conducted on four real-world datasets, including Movielens\footnote{https://movielens.org/.}, Tiktok\footnote{ https://www.tiktok.com/.}, Kwai\footnote{https://www.kwai.com/.}, and Amazon\footnote{http://jmcauley.ucsd.edu/data/amazon/.}. 
Beyond the collected user-item interaction records, these datasets provide rich content information of items. 
For fairness, we discard the feature extractor in our framework and use the feature vectors captured by some pre-trained deep learning models, including ResNet~\cite{ResNet}, VGGish~\cite{VGG}, and Sentance2Vector~\cite{S2V}, which are summarised in Table~\ref{table_1}.
\begin{table}
  \centering
  \small
  \renewcommand\arraystretch{1.0}
  \caption{Summary of the datasets. V, A, and T denote the dimensions of  feature vectors in visual, acoustic, and textual modalities, respectively.}
\vspace{-8pt}
  \label{table_1}
  \setlength{\tabcolsep}{1.0mm}
  \begin{tabular}{c|c|c|c|c|c|c|c}
    \specialrule{0.1mm}{0pt}{1pt}
    Dataset& Inter. \#& User \#& Warm \#& Cold\# & V & A & T\\
    \specialrule{0.1mm}{1pt}{1pt}
    \specialrule{0.1mm}{1pt}{1pt}
    Movielens & 922,007 & 55,485 & 5,119 & 867 & 2,048 & 128 & 100\\
    Tiktok & 389,406 & 32,309 & 57,832 & 8,624 & 256 & 128 & 128\\
    Kwai & 236,130 & 7,010 & 74,470 & 12,013 & 2,048 & - & -\\
    Amazon & 142,539 & 27,044 & 68,810 & 17,696 & 64 & - & -\\
    \hline
  \end{tabular}
  \vspace{-10pt}
\end{table}
\begin{table*}
\small
  \centering
  \renewcommand\arraystretch{1.0}
  \caption{Comparison with Cold-Start Recommendation Models over Four Datasets.}
      \vspace{-8pt}
  \begin{tabular}{c|c|ccc|ccc|ccc|ccc}
    \hline
    \multirow{2}{*}{Metric}&\multirow{2}{*}{Model}&\multicolumn{3}{c|}{Movielens}&\multicolumn{3}{c|}{Tiktok}&\multicolumn{3}{c|}{Kwai}&\multicolumn{3}{c}{Amazon}\\
    \cline{3-14}
    &&Warm&Cold&All&Warm&Cold&All&Warm&Cold&All&Warm&Cold&All\\
    \hline
    \hline
    \multirow{8}{*}{recall@10}
                        & DUIF &0.2417&\underline{0.0691}&0.1907&0.0736&\underline{0.0057}&0.0567&0.0317&\underline{0.0068}&0.0273&0.0047&\underline{0.0159}&0.0043\\
    \cline{2-14}
                        & DropoutNET &0.2437&0.0473&0.1939&0.0792&0.0034&0.0618&0.0406&0.0022&0.0349&0.0102&0.0010&0.0070\\
                        & MTPR &0.2607&0.0593&0.2059&0.0852&0.0045&0.0661&0.0464&0.0027&0.0401&0.0113&0.0028&0.0077\\
    \cline{2-14}
                        & CB2CF &\underline{0.2708}&0.0317&\underline{0.2134}&\underline{0.0870}&0.0016&\underline{0.0678}&\underline{0.0465}&0.0023&\underline{0.0403}&\underline{0.0167}&0.0139&\underline{0.0110}\\
                        & Heater &0.2523&0.0586&0.1993&0.0828&0.0032&0.0639&0.0464&0.0029&0.0400&0.0104&0.0012&0.0071\\
    \cline{2-14}
                        & CLCRec &\textbf{0.3178}&\textbf{0.0730}&\textbf{0.2485}&\textbf{0.1408}&\textbf{0.0113}&\textbf{0.1033}&\textbf{0.0648}&\textbf{0.0098}&\textbf{0.0522}&\textbf{0.0238}&\textbf{0.0313}&\textbf{0.0190}\\
                        
    \cline{2-14}
                        & \%Improv. &17.35\%&5.64\%&16.45\%&61.86\%&103.95\%&52.48\%&39.35\%&44.71\%&29.53\%&42.51\%&96.24\%&73.04\%\\
    \hline
    \hline
    \multirow{8}{*}{NDCG@10}                             
                            & DUIF &0.1577&\underline{0.0432}&0.1308&0.0426&\underline{0.0028}&0.0338&0.0273&\underline{0.0036}&0.0246&0.0021&\underline{0.0082}&0.0020\\
    \cline{2-14}
                             & DropoutNET &0.1427&0.0217&0.1191&0.0459&0.0020&0.0368&0.0303&0.0012&0.0272&0.0049&0.0004&0.0034\\
                             & MTPR &0.1720&0.0317&0.1427&0.0473&0.0025&0.0378&\underline{0.0406}&0.0017&\underline{0.0365}&0.0061&0.0014&0.0041\\
    \cline{2-14}
                             & CB2CF &\underline{0.1785}&0.0168&\underline{0.1474}&\underline{0.0503}&0.0006&\underline{0.0404}&0.0401&0.0014&0.0358&\underline{0.0088}&0.0065&\underline{0.0058}\\
                             & Heater &0.1627&0.0368&0.1349&0.0466&0.0020&0.0370&0.0397&0.0015&0.0357&0.0055&0.0006&0.0038\\
    \cline{2-14}
                            & CLCRec &\textbf{0.2392}&\textbf{0.0444}&\textbf{0.1969}&\textbf{0.0853}&\textbf{0.0062}&\textbf{0.0650}&\textbf{0.0528}&\textbf{0.0059}&\textbf{0.0451}&\textbf{0.0126}&\textbf{0.0175}&\textbf{0.0103}\\

    \cline{2-14}
                            & \%Improv. &34.01\%&2.78\%&33.85\%&69.54\%&119.01\%&61.07\%&30.05\%&64.61\%&23.56\%&42.76\%&111.89\%&77.66\%\\
    \hline
  \end{tabular}
\vspace{-6pt}
  \label{table_2}
\end{table*}
\begin{table}
\small
  \centering
  \renewcommand\arraystretch{1}
  \caption{Comparison with BPR loss-based Recommendation Models over Four Datasets.}
\vspace{-7pt}
  \label{table_3}
  \begin{tabular}{c|c|c|c|c|c}
    \hline
    \multicolumn{2}{c|}{Data}&MF-BPR&LightGCN&$CEN_{MF}$&$CEN_{GCN}$\\
    \hline
    \hline
    \multirow{3}{*}{Movielens}&Warm&0.2591&0.2136&\textbf{0.3178}&0.3110\\
                            &Cold&0.0190&0.0131&\textbf{0.0730}&0.0637\\
                            &All&0.2051&0.1942&\textbf{0.2485}&0.2441\\
    \hline
    \multirow{3}{*}{Tiktok}&Warm&0.0853&0.0941&0.1408&\textbf{0.1532}\\
                            &Cold&0.0004&0.0022&\textbf{0.0113}&0.0111\\
                            &All&0.0662&0.0729&0.1033&\textbf{0.1133}\\
    \hline
    \multirow{3}{*}{Kwai}&Warm&0.0474&0.0425&0.0648&\textbf{0.0659}\\
                        &Cold&0.0005&0.0006&0.0098&\textbf{0.0107}\\
                        &All&0.0406&0.0370&0.0522&\textbf{0.0549}\\
    \hline
    \multirow{3}{*}{Amazon}&Warm&0.0112&0.0155&0.0238&\textbf{0.0251}\\
                            &Cold&0.0008&0.0008&\textbf{0.0313}&0.0276\\
                            &All&0.0076&0.0105&0.0190&\textbf{0.0196}\\
    \hline
  \end{tabular}
\vspace{-12pt}
\end{table}
For each dataset, we randomly sample some items as the cold-start items and split them into the cold validation and testing sets with the ratio $1:1$. 
The other items are grouped into the training set, warm validation set, and warm testing set with the ratio $8:1:1$. 
And, we build the all validation (testing) set by combing the warm with cold validation (testing) sets. 
The validation and testing sets are used to tune the hyper-parameters and evaluate the performance in the experiments, respectively. 
\subsubsection{\textbf{Baselines}}
We compare CLCRec with several state-of-the-art models designed for the cold-start recommender system.  
In addition to the robustness-based (\textit{i.e.} DropoutNET, MTPR) and constraint-based methods~(\textit{i.e.} CB2CF, Heater), we introduce a content-based model~(\textit{i.e.,} DUIF) to highlight the performance of our proposed model for cold-start item recommendation. 
Furthermore, we emphasize the effectiveness of the proposed objective function by comparing two different implementations~(\textit{i.e.,} $CEN_{MF}$ and $CEN_{GCN}$) with the BPR loss-based model: MF-BPR and LightGCN.
\begin{itemize}
    \item \textbf{DUIF}~\cite{DUIF}
    Different from the CF-based model, this method learns the user reference to the item's features without modeling the CF signal, which intrinsically avoids the cold-start problem. 
    \item \textbf{DropoutNet}~\cite{DropoutNET}
    This method performs the dropout operation that randomly discards the partial collaborative embeddings to build the cold-start condition in the training phase. It alleviates the cold-start problem by improving the robustness of the model. 
    \item \textbf{MTPR}~\cite{MTPR}
    Inspired by the counterfactual thinking, this method defines counterfactual representation to replace the collaborative embedding with the all-zero vector.
    \item \textbf{CB2CF}~\cite{CB2CF}
    It performs a deep neural multiview model to represent the rich content information. And, a mean squared error~(MSE) serves as a bridge from items' feature representations into their collaborative embeddings.  
    \item \textbf{Heater}~\cite{Heater}
    This method uses the sum squared error~(SSE) loss to model the collaborative embedding from the content information. Meanwhile, it harnesses a randomized training method to promote the effectiveness. 
    \item \textbf{MF-BPR}~\cite{BPR} 
    The method learns latent vectors~(\textit{i.e.} collaborative embedding) to represent the user and item according to past interactions, and predict their affinities by measuring the similarities between the learned representations. 
    \item \textbf{LightGCN}~\cite{LightGCN}
    Based on graph convolutional networks~\cite{LZG1,LZG2}, it learns the high-order CF signal and injects it into the collaborative embedding to optimize the recommendation. 
\end{itemize}

\subsubsection{\textbf{Evaluation Metrics}}
We adopt the full ranking evaluation for the warm-start items, cold-start items, and all items, respectively. Whereinto, all items indicate the case that the warm- and cold- start item are combined to perform the experiments. 
In all cases, we adopt recall@K and Normalized Discounted Cumulative Gain~(NDCG@K) as the metrics, which are widely used in personalized recommender systems. 
By default, we set K=10 and report the average values of the above metrics for all users during the testing phase. 
\subsubsection{\textbf{Parameter Settings}}
We implement baselines and our proposed model with the help of Pytorch\footnote{https://pytorch.org/.} and torch-geometric package\footnote{https://pytorch-geometric.readthedocs.io/.}. 
Xavier~\cite{Xavier} and Adam~\cite{Adam} algorithms are utilized in the experiments to initialize and optimize the parameters of the models. 
For fairness, we set the dimension of the collaborative embedding as 64 for all models.  
In terms of the hyper-parameters, we use the grid search~\cite{JH,YXZ}: 
the learning rate is tuned in $\{0.0001, 0.001, 0.01, 0.1\}$ and regularization weight is searched in $\{0.0001, 0.001, 0.01, 0.1\}$. 
Besides, we employ the early stopping strategy~\cite{JI2}, which stops the training if recall@10 on the validation data does not increase for 10 successive epochs. 
For the baselines, we do the same options and follow the designs in their articles to achieve the best performance. 
\begin{figure}
    \centering
    \subfigure[Warm-start Item on Movielens]{
      \includegraphics[width=0.21\textwidth]{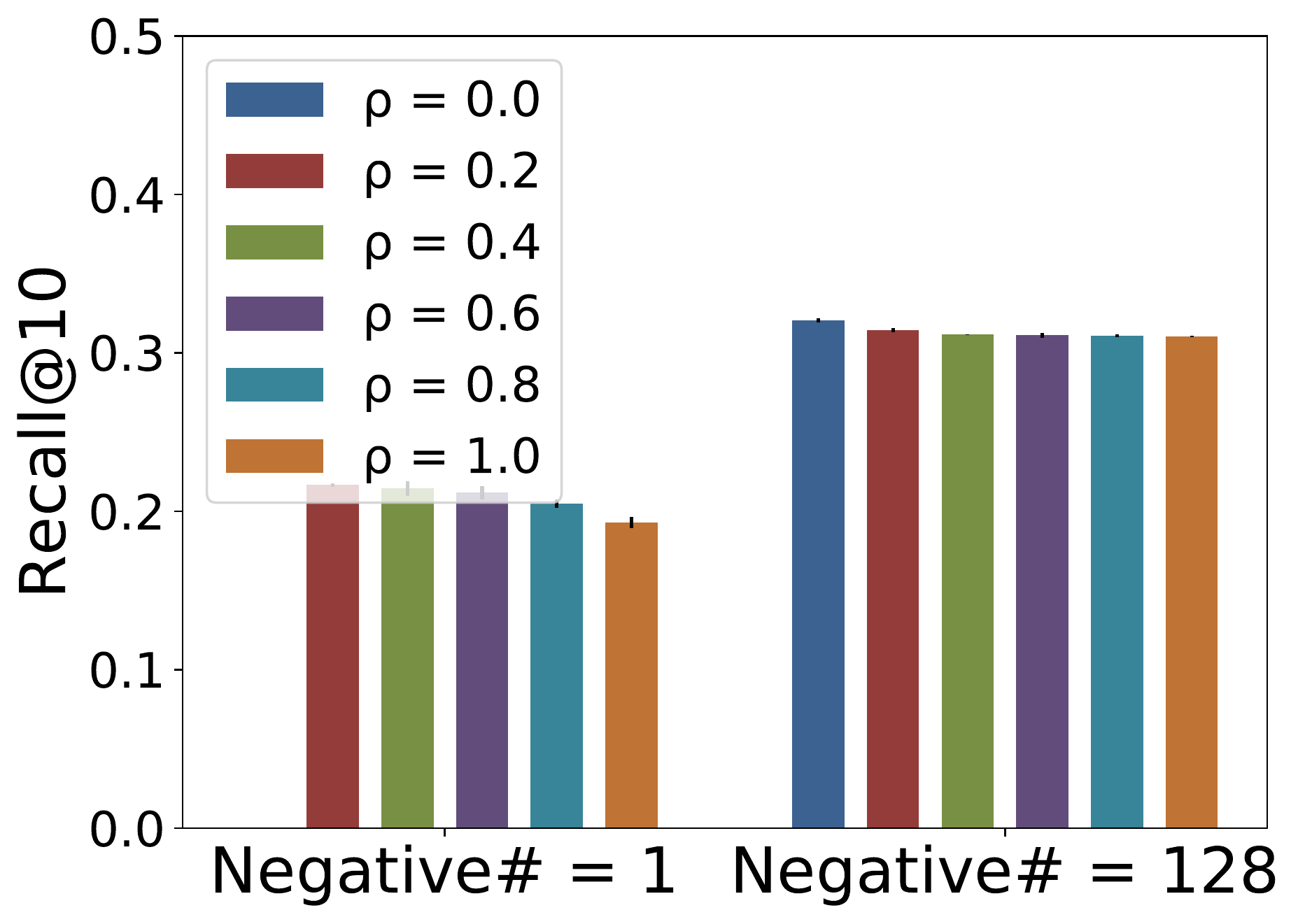}
      \label{fig_ui_mv}
    }\quad
\vspace{-5pt}
    \subfigure[Cold-start Item on Movielens]{
      \includegraphics[width=0.215\textwidth]{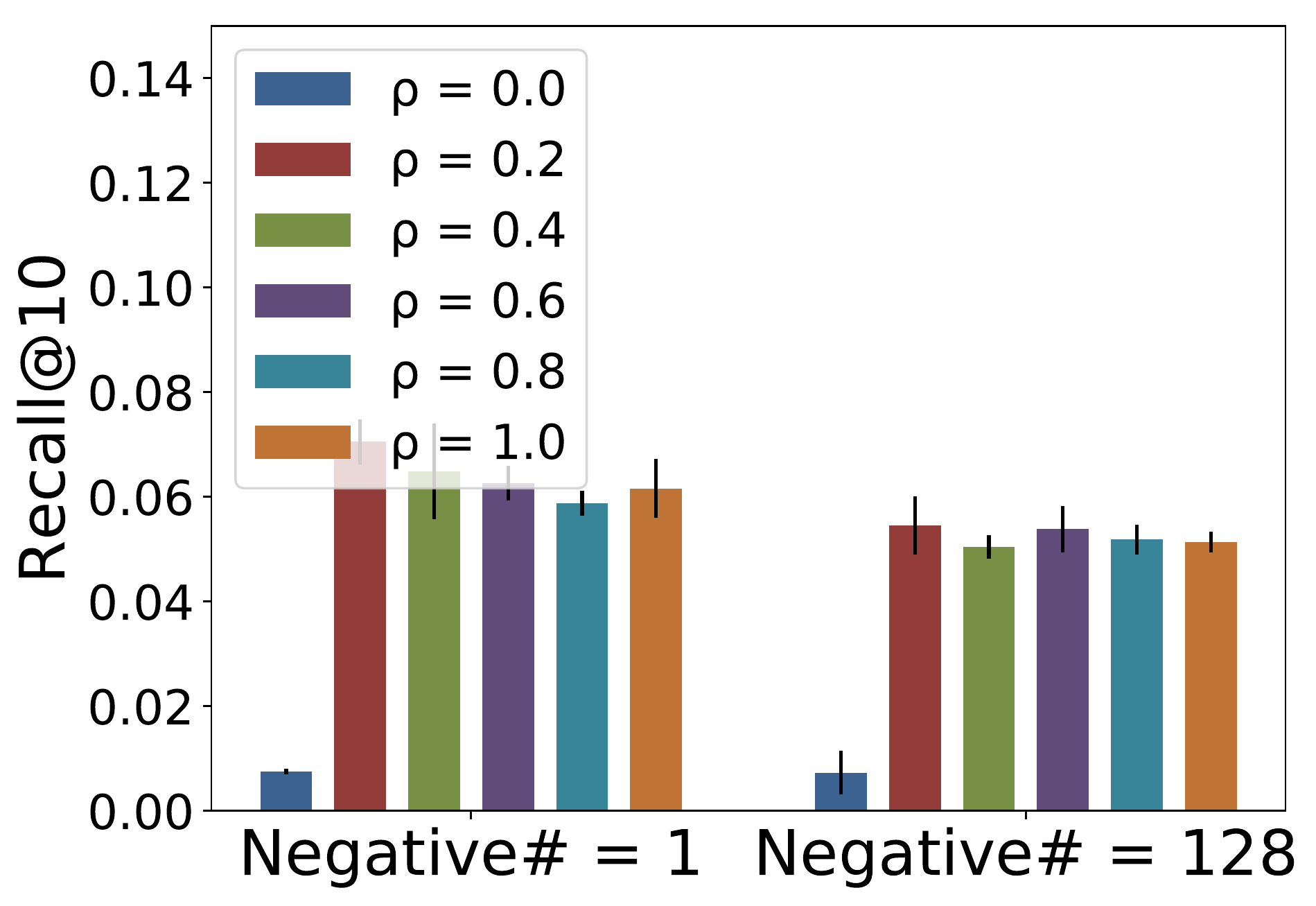}
      \label{fig_ui_mv}
    }\quad
\vspace{-5pt}
    \subfigure[Warm-start Item on Amazon]{
      \includegraphics[width=0.22\textwidth]{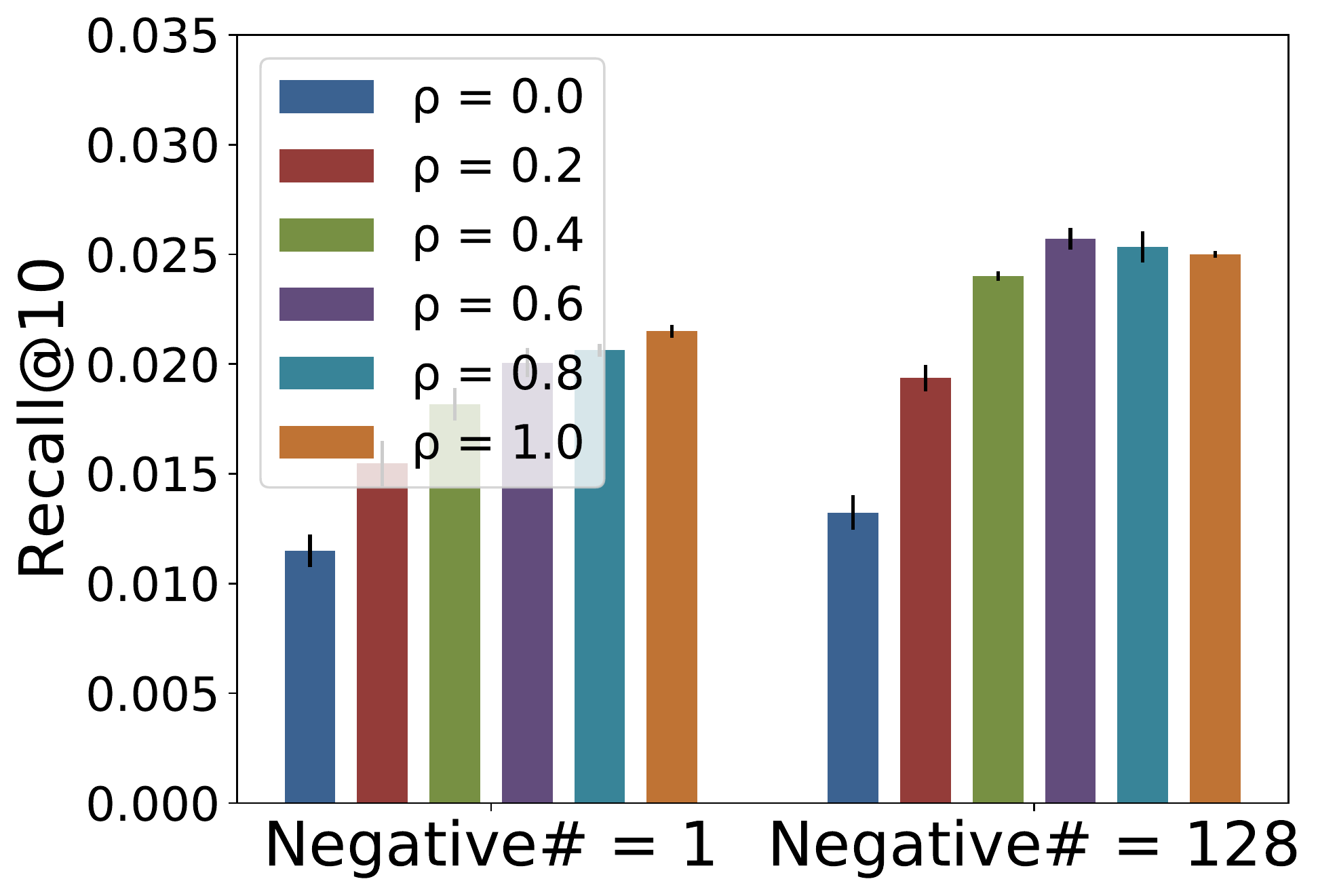}
      \label{fig_ui_amazon}
    }\quad
    \subfigure[Cold-start Item on Amazon]{
      \includegraphics[width=0.22\textwidth]{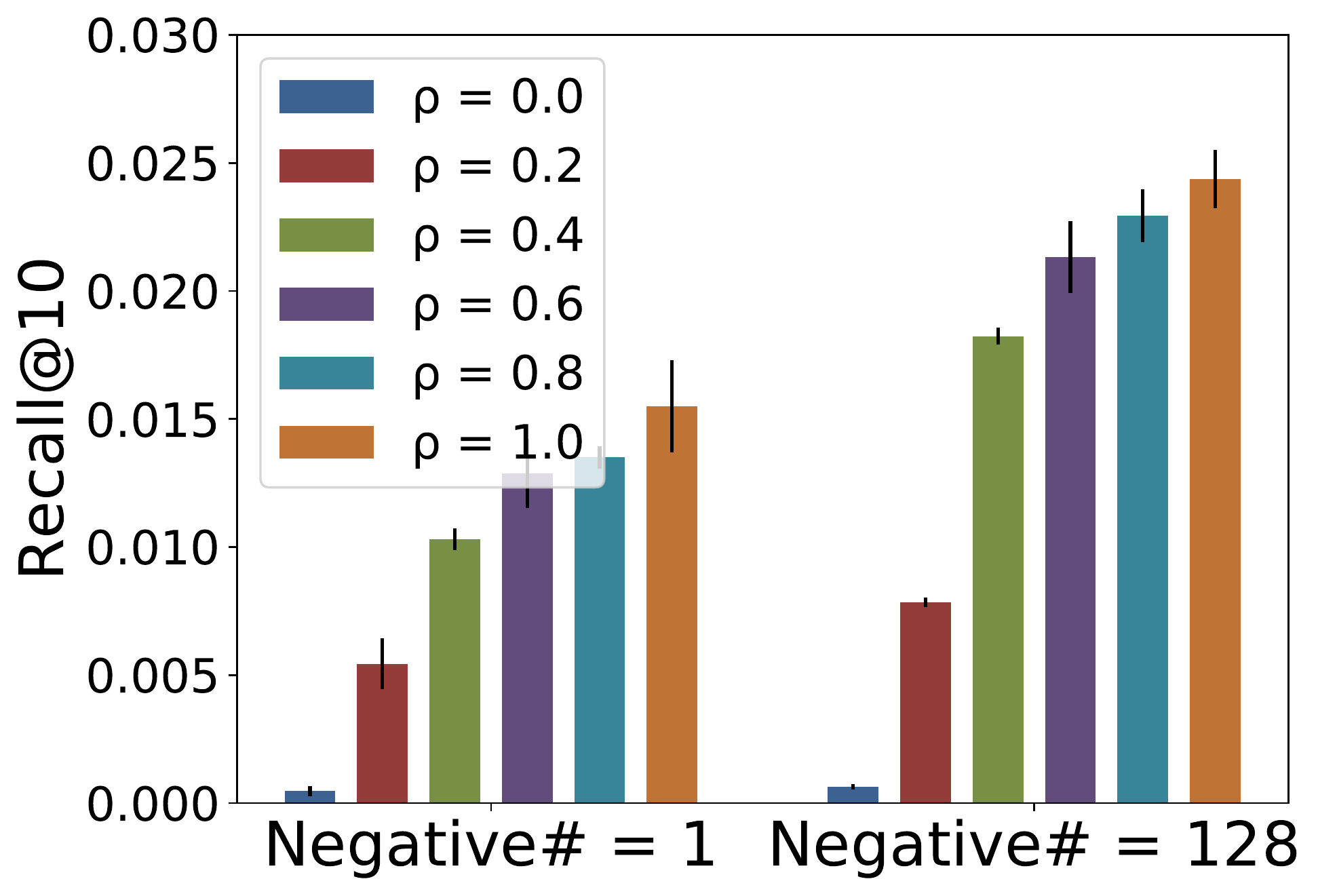}
      \label{fig_ui_amazon}
    }
\vspace{-5pt}
    \caption{Effect of U-I Contrastive Embedding Network. Negative\# denotes the number of negative pairs.}
    \label{fig_ui}
\vspace{-15pt}
 \end{figure} 
\subsection{Performance Comparison}
\subsubsection{\textbf{Empirical Results w.r.t. Cold-start Recommendation}}
Beyond the empirical results of all methods, we report the improvements in Table~\ref{table_2}, which are calculated between our proposed method and the strongest baselines highlighted with underline. 
Analyzing the table from top to bottom, we observe that:
\begin{itemize}
    \item In terms of recall@10 and NDCG@10, the performance of DUIF surpasses that of other baselines by a large margin for the items which are not observed in the training set. It is reasonable since the user in the content-based model is represented by refining the features of items she/he interacted, which associates with the unobserved item via the item features. Whereas, there is a gap between the user's collaborative embedding and item's feature representation in CF-based models. 
    \item By jointly analyzing the results in Table~\ref{table_3}, it can be found that the methods~(Dropout, MTPR, CB2CF, and Heater) designed for cold-start problem outperforms MF-BPR and LightGCN in the cold-start scenario. It indicates that these methods capture the collaborative signal from items' features, which is helpful for predicting the interactions between users and cold-start items.
    \item Undoubtedly, CLCRec outperforms the baselines in all cases~(\textit{i.e.,} warm, cold, all) across the datasets and in both evaluation metrics of recall@10 and NDCG@10. In particular, for cold-start items, CLCRec achieves the improvements over the strongest baseline \textit{w.r.t.} recall@10 by 5.64\%, 103.95\%, 44.71\%, and 96.24\% on the four datasets, respectively. In addition, compared with the other CF-based baselines, the improvements of our proposed method are more significant. It can be attributed to the following aspects: (1) CLCRec equipped with the U-I contrastive learning facilitates the collaborative embedding modeling, and (2) maximizing the R-E mutual information makes the feature representation preserve much more information related to the collaborative signal. 
    \item Analyzing the results across different datasets, we find that the improvements of our proposed method on Tiktok and Amazon are much more significant than that of Movienlens and Kwai datasets. 
    It signifies that CLCRec can distill the informative features from the content information, while the baselines are limited by the sparse interactions on Tiktok and Amazon datasets. 
\end{itemize}
\begin{figure}
    \centering
    \subfigure[Movielens]{
      \includegraphics[width=0.22\textwidth]{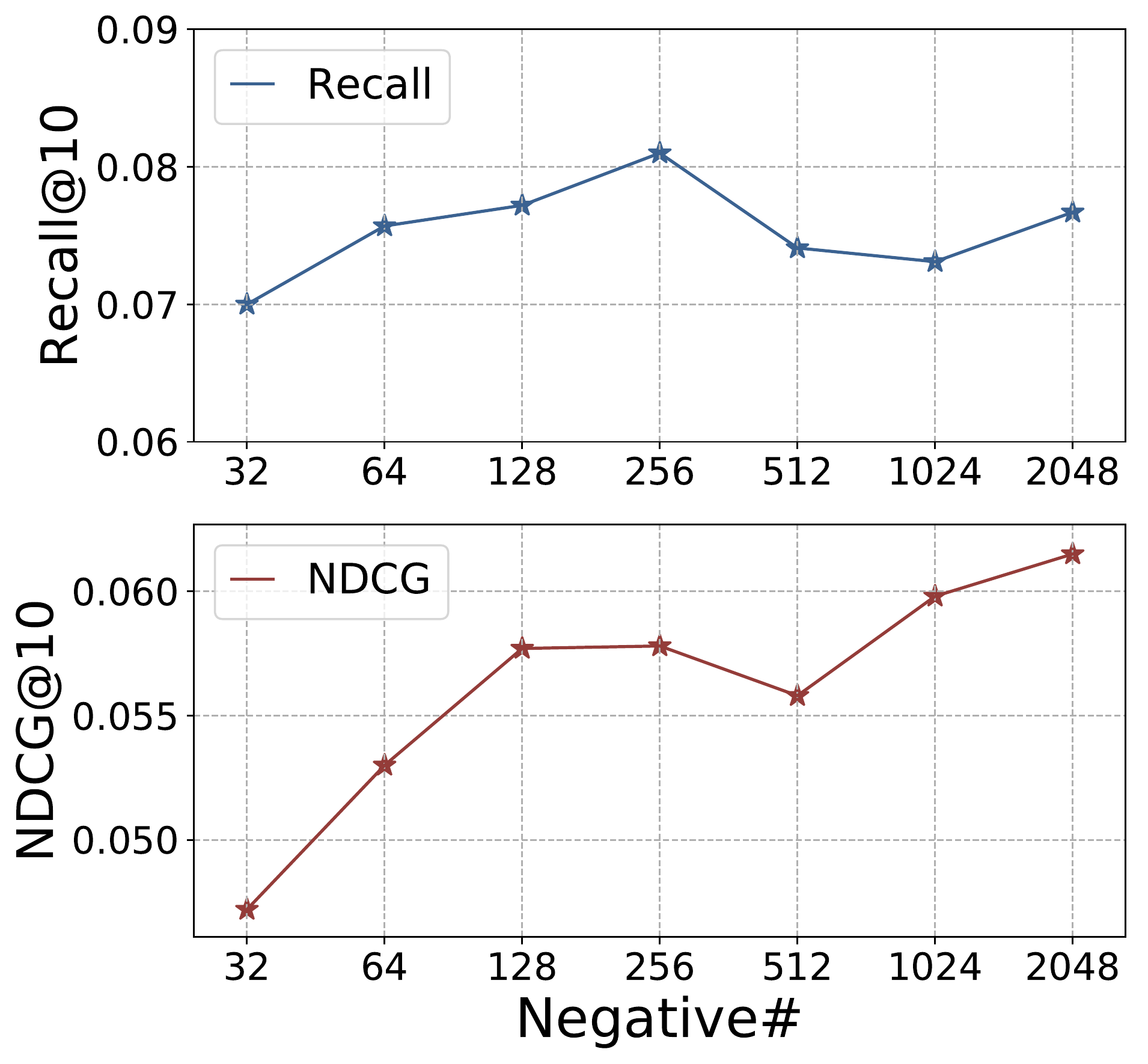}
      \label{fig_er_mv}
    }
    \subfigure[Amazon]{
      \includegraphics[width=0.22\textwidth]{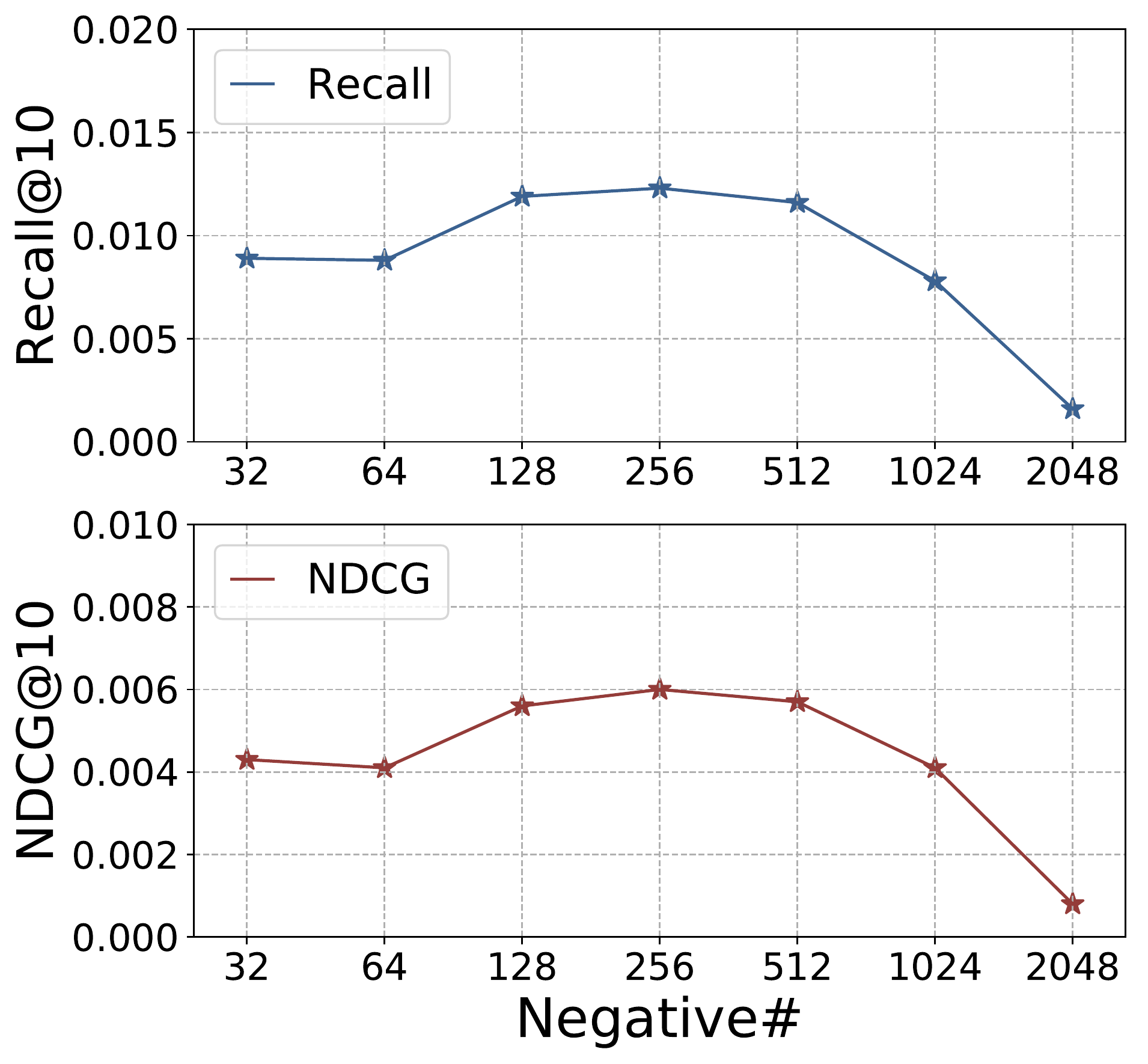}
      \label{fig_er_amazon}
    }
\vspace{-7pt}
    \caption{Effect of R-E Contrastive Embedding Network \textit{w.r.t.} recall@10 and NDCG@10. Negative\# denotes the number of negative pairs for the contrastive learning.}
    \label{fig_er}
\vspace{-7pt}
 \end{figure} 
\subsubsection{\textbf{Performance Comparison w.r.t. Objective Functions}} 
In this work, we propose an objective function founded on the contrastive learning and develop a general framework with the flexible encoders. 
To justify the objective function, we compare two different implementations equipped with MF-based and GCN-based CF encoders~($CEN_{MF}$ and $CEN_{GCN}$) with MF-BPR and LightGCN, which are optimized with BPR loss function. 
We analyze their experimental results \textit{w.r.t.} recall@10 listed in Table~\ref{table_3}:
\begin{itemize}
    \item Clearly, $CEN_{MF}$ and $CEN_{GCN}$ significantly outperform MF-BPR and LightGCN not only on the cold-start but on the warm-start conditions. It indicates that our proposed objective function is able to effectively model collaborative embedding as well as distill the features associating with the collaborative signal. 
    \item Analyzing the results on the cold-start scenario across the four datasets, we can see that $CEN_{MF}$ performs slightly better than $CEN_{GCN}$ in most cases. One possible reason is that the GCN-based model is enhanced by injecting the items' local structure information into their collaborative embeddings, while such information is hard to learn purely from their own content information by the feature encoder. 
    \item Although LightGCN could encode the local structure information into collaborative embeddings, we find the results of MF-based methods are superior on Movielens dataset. We attribute it to the dense interaction per user on this dataset, which probably results in the smoothness of collaborative embeddings and negatively influences the GCN-based recommender system. 
\end{itemize}
\subsection{Ablation Study}
In this section, we perform ablation studies to obtain deep insights on our proposed framework. 
We start by exploring U-I CEN to justify its influence towards the CF embedding and feature representation. We then investigate R-E CEN to figure out how the R-E mutual information affects the performance. In what follows, we analyze an important hyper-parameter $\lambda$, which implements the trade-off between the warm- and cold- start performance. Due to the space limitation, we omit some results which have the similar trend with the ones exhibited in the following.
\subsubsection{\textbf{Impact of U-I CEN}}
To investigate whether the proposed method can benefit from U-I CEN and the hybrid contrastive training, we discard R-E CEN and varied the hybrid sampling probability $\rho$ in the range of \{0.0, 0.2, 0.4, 0.6, 0.8, 1.0\} under single and multiple negative sampling cases, respectively. 
Figure~\ref{fig_ui} shows the results \textit{w.r.t.} recall@10 on Movielens and Amazon datasets in warm- and cold- start scenarios. 
We observe that:
\begin{itemize}
    \item Compared with the single negative pair sampling, U-I CEN with 128 negative pairs is capable of boosting the performance substantially in most cases. Focusing on the performance of warm-start items, we find that the improvements on Movielens are more significant than those on Amazon. This might be caused by the sparsity of interactions on Amazon, making the hard negative sampling easy, even for the single negative sampling.
    \item The hybrid contrastive training serves as \textit{vitamin} to the cold-start recommendation. That is, even a small $\rho$, like 0.2, is beneficial to the feature representation learning, which verifies the effectiveness of this training strategy. 
    \item From the results, we observe that the performance on Movielens dataset slightly decreases when sampling multiple negative pairs and increasing the sampling probability $\rho$. It is probably caused by the densely interacted items on Movielens, which makes the information unrelated to collaborative signal accumulate. With the increasing negative samples and $\rho$, much more noisy information is encoded into the collaborative embedding. 
\end{itemize}
\begin{figure}
    \vspace{-5pt}
    \centering
    \subfigure[recall@10 on Movielens]{
      \includegraphics[width=0.21\textwidth]{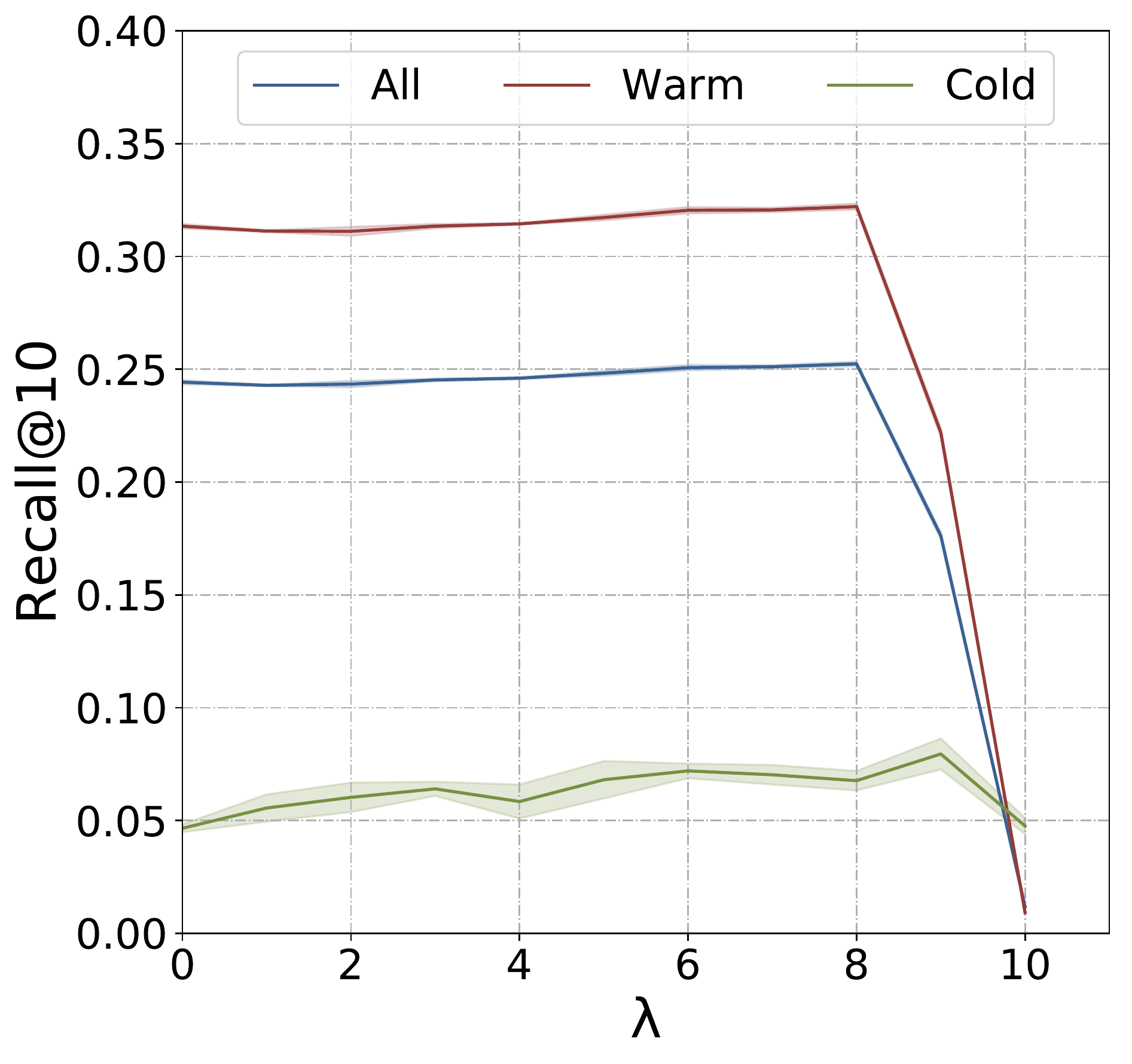}
    }
    \subfigure[recall@10 on Amazon]{
      \includegraphics[width=0.215\textwidth]{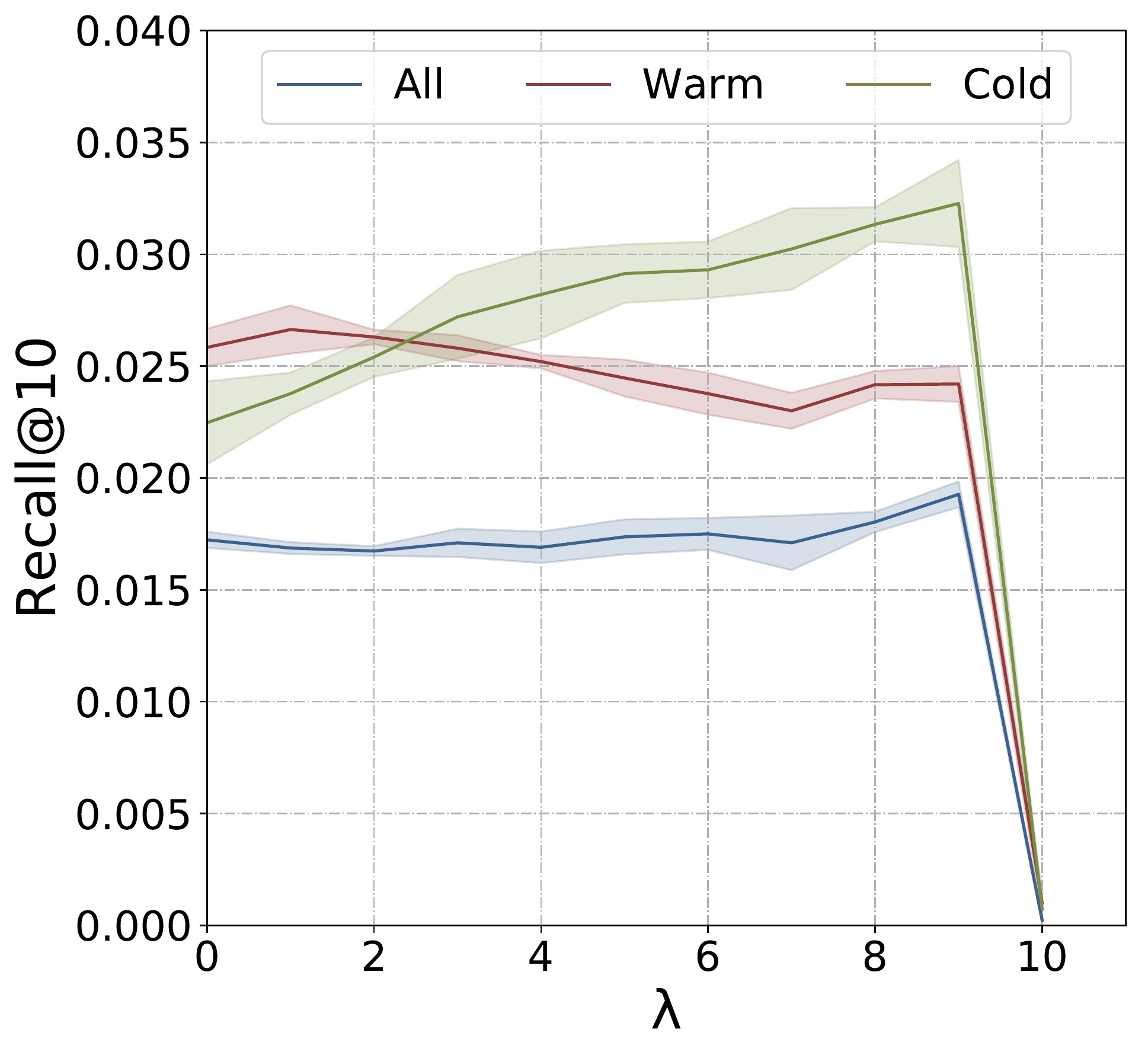}
    }
\vspace{-10pt}
    \caption{Performance comparison \textit{w.r.t.} different $\lambda$ on Movielens and Amazon.}
    \label{fig_lambda}
\vspace{-12pt}
 \end{figure} 
\subsubsection{\textbf{Impact of R-E CEN}}
To demonstrate the effectiveness of R-E CEN, we do ablation study by adjusting the number of R-E negative pairs. In particular, we sample one negative user-item pair for U-I CEN to emphasize the influence of R-E CEN for the performance of cold-start items. The results \textit{w.r.t.} recall@10 and NDCG@10 on Movielens and Amazon datasets are plotted in Figure~\ref{fig_er}, from which we derive two key observations:
\begin{itemize}
    \item Jointly analyzing the results in Table~\ref{table_2}, our proposed method shows significant improvements over all competing methods even when we only sample several negative pairs for contrastive training. Moreover, increasing the negative pair number from 32 to 256 significantly enhances the performance. Taking Movielens dataset as an example, setting the number as 256 leads to the highest recall@10 of 0.0810. It illustrates the substantial influence of R-E CEN on item recommendation.
    \item However, the performance drops when the number of negative pairs goes beyond 256, especially for the Amazon dataset. One possible reason is that the overload negative sampling causes the overfitting of R-E mutual information, which corrupts the expressiveness of the collaborative embedding. 
    This suggests that, although our proposed method benefits from the multiple negative pairs, it also could be suffered from the overuse.
\end{itemize}

\subsubsection{\textbf{Trade-off between Warm- and Cold- Start}}
We vary the quantity of the trade-off hyper-parameter~(\textit{i.e.,} $\lambda$) to investigate the effectiveness of U-I and R-E CENs towards the warm- and cold- start item recommendation. 
Observing the performance shown in Figure~\ref{fig_lambda}, we have the following findings:
\begin{itemize}
    \item It is not unexpected that the performance \textit{w.r.t.} recall@10 consistently improves in the cold-start case, when increasing the value of $\lambda$ in range of \{0.0, 0.1, \dots, 0.8\}. Despite the visually smooth curve on Movielens dataset, the absolute value of recall@10 increases from 0.0463 at the beginning to the highest value 0.0730. It demonstrates that $\mathcal{L}_{RE}$ is beneficial to capture the collaborative information from the content for the cold-start recommendation. 
    \item For the warm-start items, the performance is slightly affected by varying $\lambda$ in most cases. It again verifies that the improvements of the cold-start recommendation are mainly obtained by maximizing the mutual information between items' collaborative embeddings and feature representations. The contribution of U-I mutual information is to enrich the collaborative embedding.
    \item Although the content information could be supplementary for collaborative signal, the performance of the warm-start items decreases with the increase of contribution of R-E mutual information, especially for the Amazon dataset. The main reason is that more information unrelated to the collaborative signal is injected into the collaborative embedding by the back-propagation during the training. From the results for the all-item scenario, it demonstrates that an appropriate value for $\lambda$ is essential to balance these two kinds of mutual information for pursuing the best performance on all of scenarios. 
\end{itemize}
\begin{figure}
    \centering
    \vspace{-5pt}
    \subfigure[gradient magnitude on Movielens]{
      \includegraphics[width=0.22\textwidth]{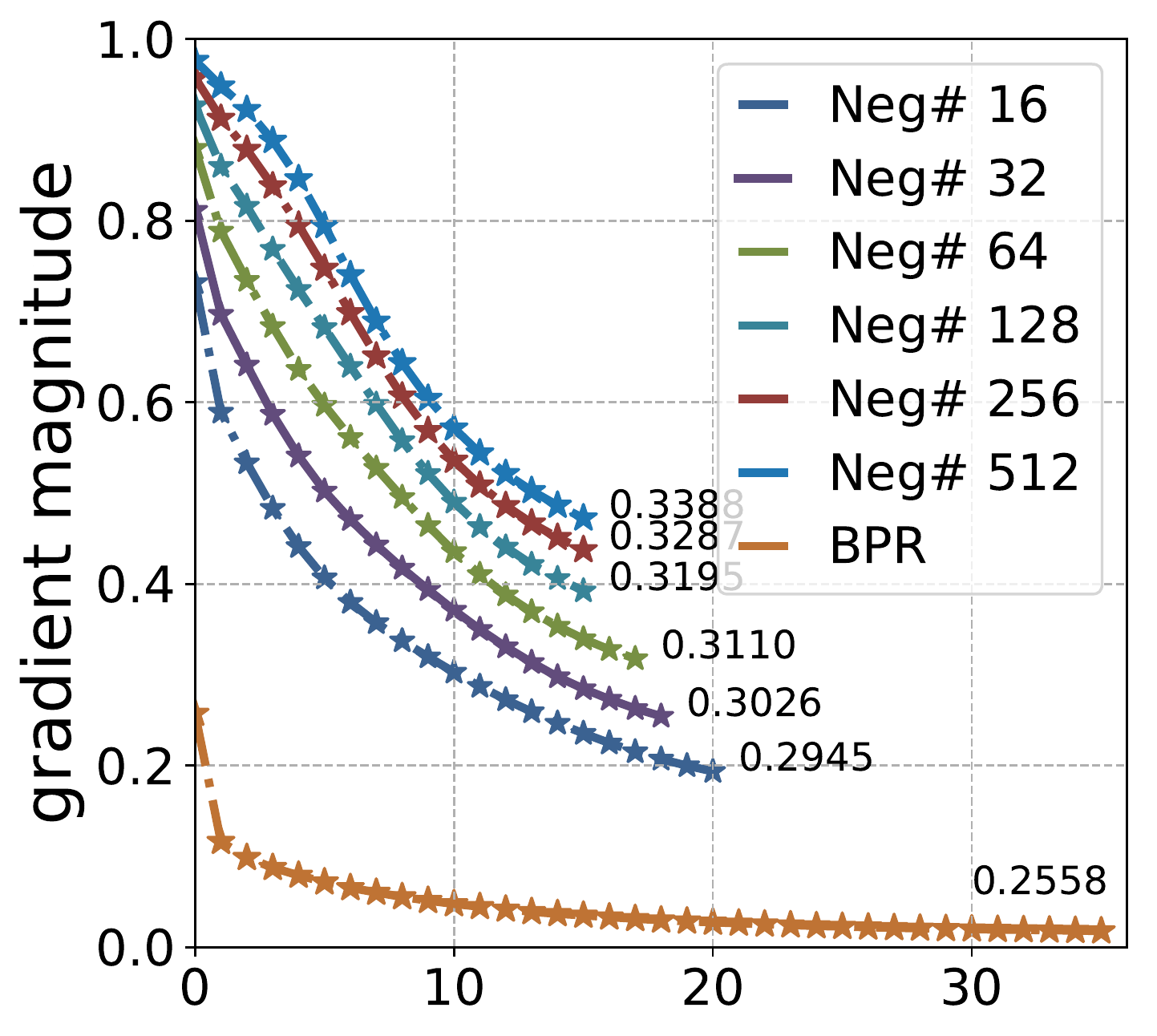}
      \label{fig_mg_mv}
    }
    \subfigure[gradient magnitude on Amazon]{
      \includegraphics[width=0.22\textwidth]{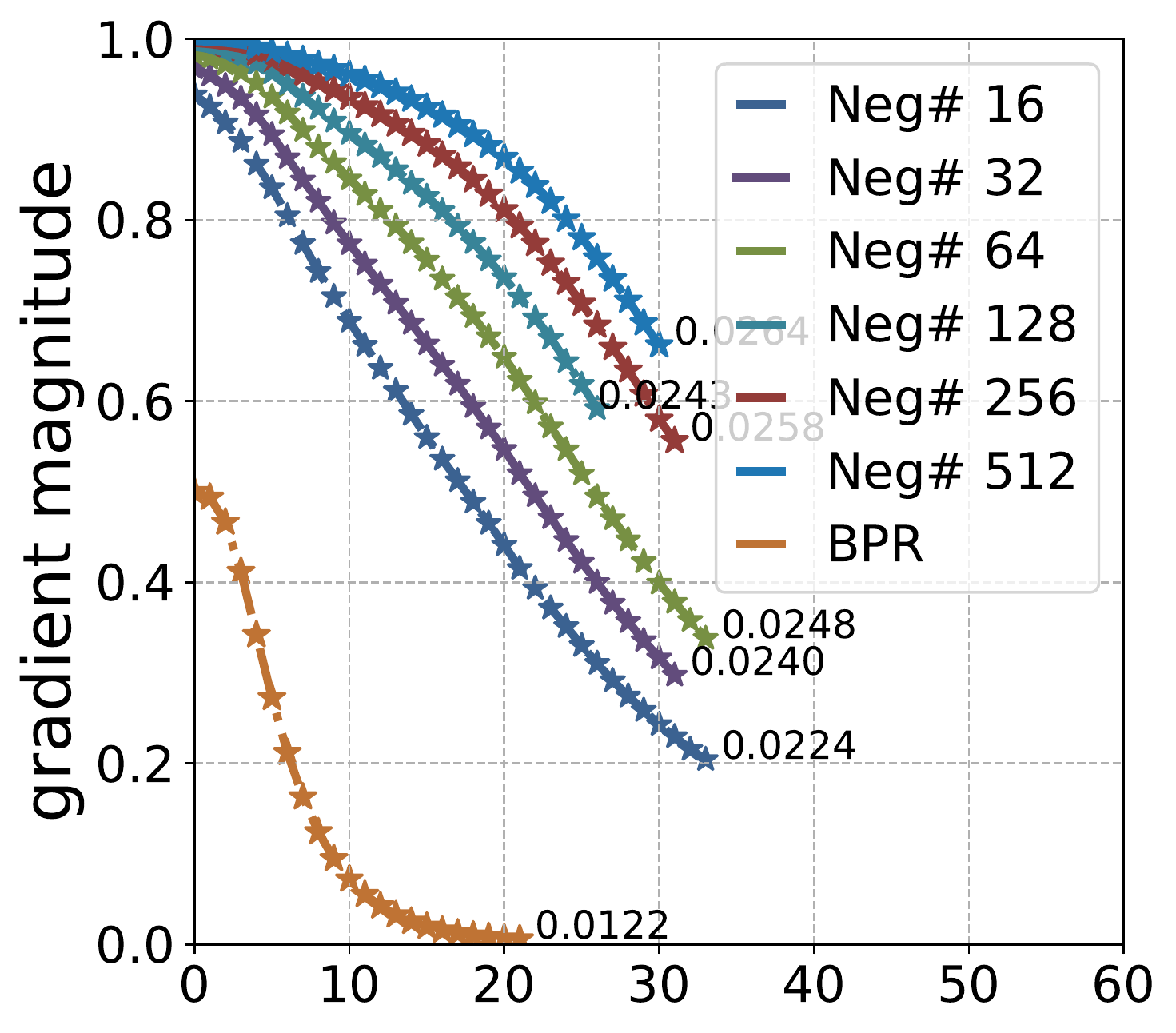}
      \label{fig_lambda_mv}
    }
\vspace{-5pt}
    \caption{Training process \textit{w.r.t.} gradient magnitude on Movielens and Amazon datasets. Neg\# denotes the number of negative pairs.}
    \label{fig_mg}
\vspace{-10pt}
 \end{figure} 
\subsection{In-depth Analysis}
To further explore how our proposed method optimizes the recommendation, we conduct the additional experiments to investigate the training process w.r.t. gradient magnitude.  Following the prior work~\cite{KGPolicy}, we test the proposed method in different numbers of contrastive negative pairs and compute the gradient magnitude in Eq.~\ref{eq_mg} as the evaluation metric. We record the status of gradient magnitude at each epoch and exhibit the learning curves associated with the performance \textit{w.r.t.} recall@10 on Movielens and Amazon datasets in Figure~\ref{fig_mg}. 

Through analyzing the average gradient magnitude and the performance, we find that our proposed method achieves the larger gradient magnitudes as compared to MF-BPR; moreover, the performance is boosting with increasing the number of negative pairs. Furthermore, the contrastive learning not only improves the recommendation performance but promotes the convergence on Movielens dataset. We attribute its efficiency to the intrinsic hard negative detection. This demonstrates that the optimization is benefited from the contrastive training strategy, which prevents the gradient vanishing problem during the training process. 
\section{Related Work}
\subsection{Cold-start Recommendation}
Towards solving the cold-start problem, it is crucial to represent the item to alleviate the difference between the warm- and cold- start items.  
Thus, the side information, especially the content features,  tends to be incorporated into the CF-based recommendation models~\cite{DJ1,DJ2}, as the bridge to model the collaborative signal for cold-start items. 
For instance, DropoutNET~\cite{DropoutNET}, MTPR~\cite{MTPR}, and CC-CC~\cite{CCCC} randomly forgo some collaborative embeddings to improve the robustness of CF-based model, which implicitly discovers the information related to collaborative signal from the items' content features. 
Different from these robustness-based models, some efforts~\cite{CB2CF,Heater,LSI,LLAE} are dedicated to explicitly modeling the correlation between the content information and collaborative embeddings. By devising the various constraint functions, such as local geometric similarity and mean square error, they enforce the representation learned from content information to approximate the collaborative embeddings. 
By contrast, we explicitly model the correlation between the content information and collaborative filtering signal by maximizing the mutual information instead of imposing strict constraint loss.   
\subsection{Contrastive Learning}
The contrastive learning, as a representative algorithm of the self-supervised learning, is widely used in the field of computer vision~\cite{SCL,SimCLR,BYOL,Moco} and natural language processing~\cite{CPC,Infoxlm}. By identifying the positive pair from some negative ones, it maximizes the mutual information between different representations, which discovers the semantic information shared by different views. 
Towards this end, Oord~\textit{et al.}~\cite{CPC} proposed a probabilistic contrastive loss, termed InfoNCE, in a way that maximally preserves the mutual information between the observation and context signals. 
Proposed independently of CPC, Hjelm~\textit{et al.}~\cite{DIM} compared the global and local features of each image and maximized their mutual information to discover the invariant semantic signal. With their success on representation learning, several approaches are proposed for various applications. 
For the recommender system, several models~\cite{ssl2} employ the contrastive learning to optimize the representations of users or items. However, the methods focus on the representation from  either the collaborative or content space. In contrast, we treat the collaborative embedding and feature representation as two different views and maximize their mutual information, in order to capture the collaborative signal from the content information. 
\section{Conclusion and Future Work}
In this work, we focus on the cold-start problem, especially for the complete cold-start items. 
By exploring the disadvantages of prior works, we propose to reformulate the cold-start item representation learning and prove it to be lower-bounded by the integration of U-I and R-E mutual information. 
And, we develop a novel objective function founded on the contrastive learning to maximize these two kinds of mutual information. 
Accordingly, we further develop a general cold-start recommendation framework, named Contrastive Learning-based Cold-start Recommendation, consisting of the contrastive pair organization, contrastive embedding network, and contrastive optimization. 
We conduct extensive experiments over four datasets, and verify the effectiveness and efficiency of our proposed method, which outperforms the state-of-the-art baselines by a large margin both in the warm- and cold- start scenarios.

To the best of our knowledge, this is the first attempt to solve the cold-start problem by maximizing the mutual information between feature representations and collaborative embeddings of items. 
In future, we plan to explore the related tasks in mutual information in recommendation, such as the explainable recommender system through disentangling the mixture information of users' and items' representations. 
Besides, we expect to further investigate the effectiveness of mutual information maximization in various cross-modal applications, like the video localization~\cite{LM1} and fashion matching~\cite{ZN,YX1,YX2}. 
\bibliographystyle{ACM-Reference-Format}
\balance
\bibliography{reference}

\end{document}